\documentclass[
 reprint,
 aps,
 prl,
 superscriptaddress
]{revtex4-1}
\usepackage{hyperref}
\usepackage{amsmath}
\usepackage{amssymb}
\usepackage[pdftex]{graphicx}
\usepackage{microtype}
\usepackage{verbatim}
\usepackage{todonotes}
\presetkeys{todonotes}{inline,backgroundcolor=yellow}{}
\usepackage{enumitem}
\usepackage{caption}

\makeatletter
\DeclareRobustCommand{\element}[1]{\@element#1\@nil}
\def\@element#1#2\@nil{%
  #1%
  \if\relax#2\relax\else\MakeLowercase{#2}\fi}
\makeatother

\newcommand{\mrm}{\mathrm}

\newcommand{\ket}[1]{{\left| {#1} \right\rangle}}
\newcommand{\bra}[1]{{\left\langle {#1} \right|}}

\newcommand{\avg}[1]{\left\langle {#1} \right\rangle }

\newcommand{\Evec}{\vec{\mathcal{E}}}

\newcommand{\Bvec}{\vec{\mathcal{B}}}
\newcommand{\Esca}{\mathcal{E}}

\newcommand{\Bsca}{\mathcal{B}}

\begin{document}
\title{Electron electric dipole moment searches using clock transitions in ultracold molecules}
\author{Mohit Verma}
\affiliation{Department of Physics, University of Toronto,  Canada M5S 1A7}
\author{Andrew M. Jayich}
\affiliation{Department of Physics, University of California Santa Barbara, USA 93106}
\author{Amar C. Vutha}
\email{amar.vutha@utoronto.ca}
\affiliation{Department of Physics, University of Toronto,  Canada M5S 1A7}

\begin{abstract}
Permanent electric dipole moments (EDMs) of fundamental particles such as the electron are signatures of parity and time-reversal violation due to physics beyond the standard model. EDM measurements probe new physics at energy scales well beyond the reach of present-day colliders. Recent advances in assembling molecules from ultracold atoms have opened up new opportunities for improving the reach of EDM experiments. But better measurement techniques, that are not limited by the magnetic field sensitivity of such molecules, are necessary before these opportunities can be fully exploited. We present a technique that takes advantage of magnetically-insensitive hyperfine clock transitions in polar molecules, and offers new ways to improve both the precision and accuracy of EDM searches with ultracold assembled molecules. 

\end{abstract}
\maketitle

Polar molecules offer one of the best ways to probe the unknown physics that led to the imbalance between matter and anti-matter in the universe \cite{DeMille2017,Safronova2018}. Precise measurements using heavy polar molecules, wherein electron spins experience enormous relativistic electric fields, have set stringent bounds on the parity ($P$) and time-reversal ($T$) violating permanent electric dipole moment (EDM) of the electron \cite{Hudson2011,Baron2014,Cairncross2017,Andreev2018} -- such experiments constrain the parameter space of new physics models out to energy scales exceeding 10 TeV \cite{Andreev2018,Feng2013}. %Polar molecules have also been used to constrain new physics by placing bounds on $P,T$-violation in nuclei \cite{Cho1989}. 

Advances in producing cold molecules \cite{Julienne2018}, such as direct laser-cooling of polar molecules \cite{Truppe2017,Collopy2018,Cheuk2018} and ultracold assembly of molecules from atoms \cite{Molony2014,Takekoshi2014,Park2015,Guo2016,Rvachov2017,Barbe2018,Guttridge2018,Green2019,Hu2019,DeMarco2019}, have generated interest in applying these techniques to EDM searches \cite{Sunaga2018,Lim2018,Fleig_DeMille,Norrgard2017,Kozyryev2017}. Large ensembles of trapped polar molecules can potentially improve  experimental sensitivity to $P,T$-violating physics by more than two orders of magnitude, due to the long trap lifetimes ($>$ 10 s) that can be realized. However, it continues to be difficult to directly laser-cool EDM-sensitive molecules -- which are typically heavier and have more numerous leakage channels out of the cooling cycle -- to the ultracold ($\lesssim$ 10 $\mu$K) temperatures needed to confine them in optical traps \cite{Julienne2018}. In this context therefore, an especially attractive and feasible path to producing ultracold trapped molecules is to assemble them from ultracold trapped atoms. A variety of ultracold polar diatomics (typically bialkali and alkali-alkaline-earth molecules) have been produced in this way (\cite{Molony2014,Takekoshi2014,Park2015,Guo2016,Rvachov2017,Barbe2018,Guttridge2018,Green2019,Hu2019,DeMarco2019}), and excellent coherence times for their hyperfine states have been demonstrated \cite{Park2017}.

However, an important challenge needs to be overcome before a sensitive EDM experiment with ultracold assembled molecules can be realized. Electron EDM measurements require molecules with unpaired electron spins, which rules out bialkali molecules in their ground states, and thus we are left with molecules with one valence electron such as YbAg (described further below). But the simple $^2\Sigma$ electronic ground states of these molecules pose problems for traditional EDM measurements: the coherence time of spin precession measurements is degraded by magnetic field noise, and they are susceptible to systematic errors from spurious magnetic fields. Both these disadvantages can be traced back to the relatively large magnetic moment of the unpaired electron spin in these simple diatomic molecules. Therefore it is believed that ultracold assembled molecules are not ideal for EDM searches, and other molecules with more complex level structures must instead be used (cf. \cite{Kozyryev2017}). 

To address this challenge, we present an EDM measurement technique which can be used with any polar molecule that has magnetically-insensitive hyperfine states (``clock states''). Such states are generically found across a host of polar molecules, including many examples of molecules that can be assembled from ultracold atoms \cite{SM}. Importantly therefore, our technique unlocks the full potential of ultracold assembled molecules for precise EDM measurements with long coherence times, in aid of the search for new physics at the $\sim$100-TeV energy scale.

We illustrate the features of our technique using the example molecule $^{174}$Yb$^{107}$Ag. This molecule belongs to a class of electron-EDM-sensitive diatomics whose constituent atoms can be laser-cooled and trapped. We anticipate that YbAg can be produced and trapped in significant quantities at ultracold temperatures in an optical trap, after assembly from ultracold Yb and Ag atoms. 
%Both the atomic precursors have been laser-cooled and trapped \cite{Yb_trapping_ref,Uhlenberg2000}. 
YbAg molecules can be synthesized at ultracold temperatures using methods similar to those demonstrated for other isoelectronic molecules (YbLi \cite{Green2019} and YbCs \cite{Guttridge2018}). We focus on YbAg rather than a Yb-alkali molecule \cite{Meyer2009} due to the larger electronegativity of Ag compared to the alkali atoms, which results in a more strongly polar molecule with enhanced sensitivity to the electron EDM \cite{Fleig_DeMille}.  %, RbHg \cite{Borkowski2017}). %The effective electric field, $\Esca_\mrm{eff}$, in YbAg is expected to be comparable to the large field of YbF \cite{Fleig}. 
In the $^2\Sigma$ electronic ground state of YbAg, the lowest rovibrational manifold contains four hyperfine states from coupling the valence electron spin ($S=1/2$) to the Ag nuclear spin ($I=1/2$). The interaction Hamiltonian for these states with external electric and magnetic fields is
\begin{equation}\label{eq:hint_1}
H_I = -\mu_B \left( g_s \vec{S} + g_I \vec{I} \right) \cdot \Bvec - D \, \hat{n} \cdot \Evec + W_{PT} \, \vec{S} \cdot \hat{n},
\end{equation}
(with $\hbar=1$ everywhere), $g_S, g_I$ are the electron and nuclear spin g-factors, $D$ is the molecular dipole moment, and $\hat{n}$ is the unit vector pointing along the internuclear axis of the molecule. The $P$,$T$-violating physics is described by the effective low-energy Hamiltonian $W_{PT} \vec{S}\cdot \hat{n}$. %The scalar $W_{PT}$ is the sum of two $P,T$-violating terms, $W_{PT} = W_d + W_{SP}$, which are expectation values of the relativistic electron EDM interaction with the electric field inside a fully oriented polar molecule ($W_d$), and the $P,T$-violating scalar-pseudoscalar interaction between the electron and nucleus ($W_{SP}$), in the molecular electronic state (cf.\ \cite{Commins1999}). %The quantity $W_d$ is sometimes written as $d_e \Esca_\mrm{eff}$ in the literature. <- I commented this out as just simply too detailed.
%where the goal of EDM searches is to set limits on the size of $W_{PT}$ \cite{DeMille2017,Safronova2018}. 

In electron EDM measurements, a lab electric field, $\Evec = \Esca_z \hat{z}$, polarizes the molecule and a small magnetic field, $\Bvec = \Bsca_z \hat{z}$, is used to control the electron spin. A molecule polarized in an electric field has a nonzero expectation value of its orientation, $\zeta = \avg{\hat{n} \cdot \hat{z}}$, and the effective interaction Hamiltonian for the electron and nuclear spin degrees of freedom can be expressed as
\begin{equation}\label{eq:hint_2}
H_\mrm{eff} = -\left( g_s S_z + g_I I_z \right) \mu_B \Bsca_z + W_{PT} S_z \, \zeta.
\end{equation}
The dependence of the molecular orientation $\zeta$ on the applied electric field $\Esca_z$ is discussed in detail in the Supplemental Material \cite{SM}, using both a detailed numerical model and a simple analytical model.  

\begin{figure}[h!]
    \centering
    \includegraphics[width=1.05\columnwidth]{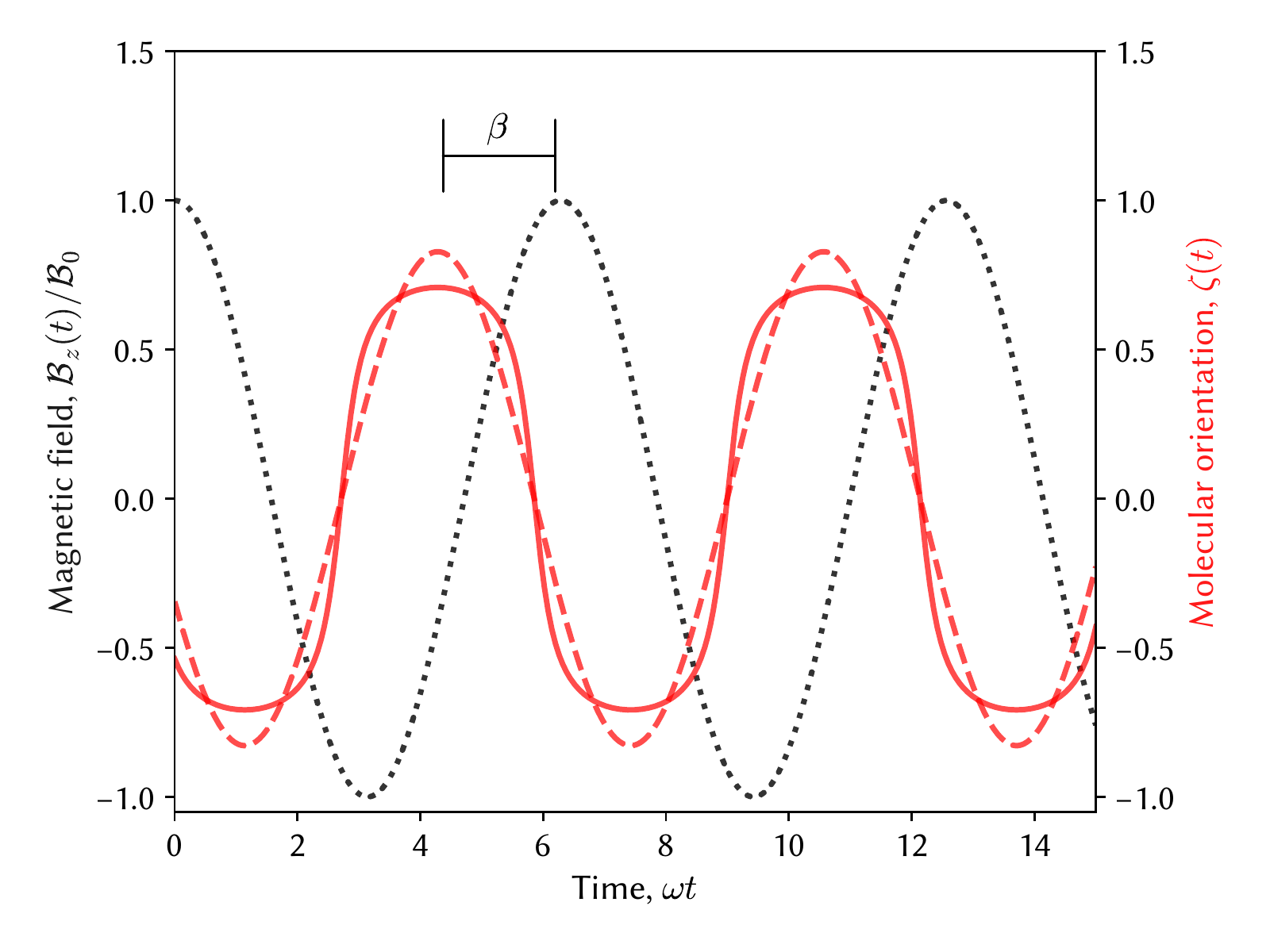}
    \caption{Magnetic field $\Bsca_z(t)$ (black, dots), and molecular orientation $\zeta(t)$ (red, solid). The curve for $\zeta(t)$ is in response to an electric field $\Esca_z(t) = \Esca_0 \cos(\omega t + \beta)$, and is calculated using the methods in (\cite{SM}, Sec.\ A). The nonlinear response of $\zeta$ to $\Esca_z$ is evident. The red dashed line shows the first harmonic of $\omega$ contained in $\zeta(t)$, which drives the hyperfine transition.}
    \label{fig:Bfield_zeta_vs_time}
\end{figure}

\begin{figure}[h!]
\centering
    \includegraphics[width=0.63\columnwidth]{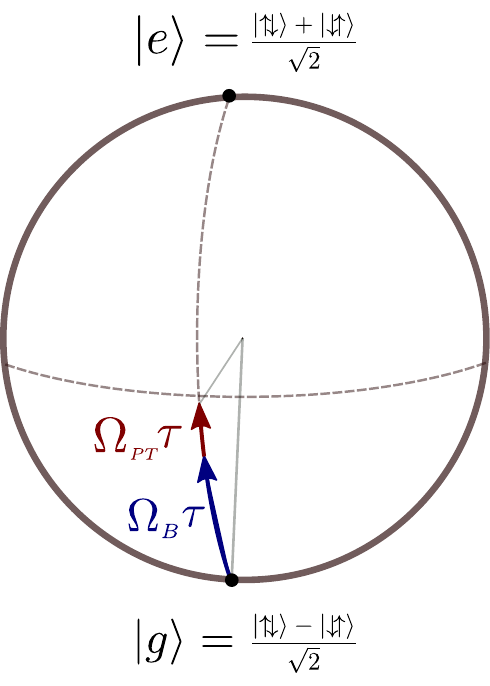}
\caption{The population transfer is shown on a Bloch sphere for the two clock states in the presence of oscillating electric and magnetic fields. The $P,T$-violating Hamiltonian leads to an extra transition amplitude $\Omega_{PT} \tau$ that interferes with the transition amplitude $\Omega_B \tau$ due to the oscillating magnetic field. The case shown here corresponds to $\beta=0$, where these amplitudes add constructively.}
\label{fig:bloch_sphere}
\end{figure}

We focus on the two hyperfine clock states $\ket{g} \equiv \ket{F=0,m_F=0}$ and $\ket{e} \equiv \ket{F=1,m_{F}=0}$, which are separated in energy by $\omega_0$. Despite the fact that both these states have zero spin (and thus zero magnetic moment), they can still be used to measure the EDM associated with the electron spin, as we demonstrate below. We propose applying a time-dependent polarizing electric field, $\Esca_z = \Esca_0 \cos(\omega_E t + \beta)$, and a time-dependent magnetic field, $\Bsca_z = \Bsca_0 \cos (\omega_B t)$, where $\beta$ is an adjustable phase. The magnetic field drives the hyperfine clock transition between $\ket{g}$ and $\ket{e}$. The electric field induces an oscillating molecular orientation with amplitude $\zeta_0$ at the frequency $\omega_E$, as shown in Fig.\ \ref{fig:Bfield_zeta_vs_time}. Due to $P,T$-violation, the molecular orientation behaves like an effective magnetic field coupled to the electron spin (see Eq. \ref{eq:hint_2}). The key idea is that the $P,T$-violating term induces an extra transition amplitude between $\ket{g}$ and $\ket{e}$, which interferes constructively or destructively (depending on the phase $\beta$) with the transition amplitude due to the applied magnetic field. The dynamics in the subspace spanned by the clock states is graphically represented on the Bloch sphere shown in Fig. \ref{fig:bloch_sphere}, which illustrates this interference between the transition amplitudes. 

\begin{figure}[h!]
\centering
    \includegraphics[width=\columnwidth]{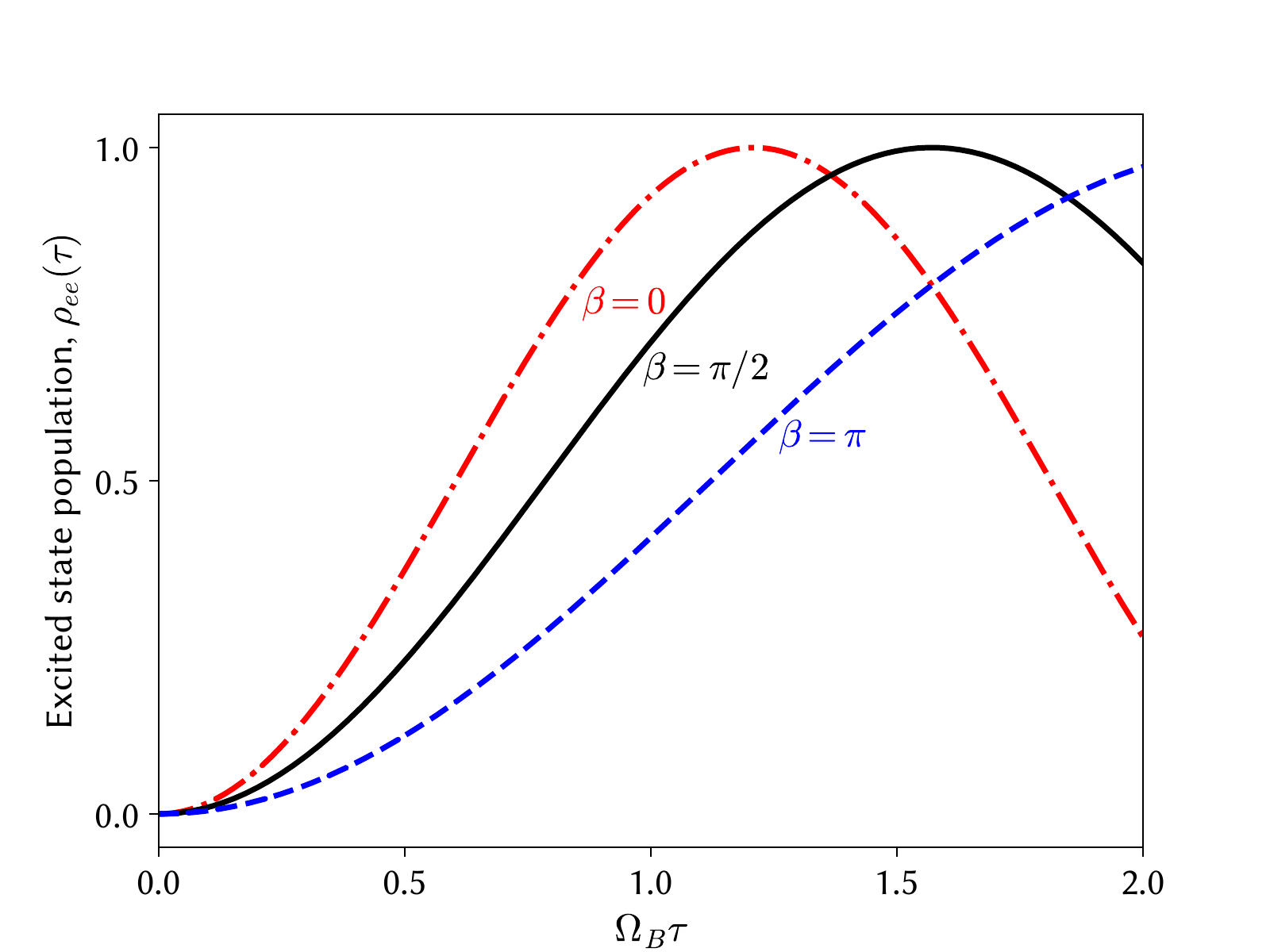}
\caption{The population in the excited clock state, $\ket{e}$, as a function of time, when the electric and magnetic fields are on resonance ($\Omega_{PT}$ is exaggerated for illustration). The relative phase $\beta$ between the electric and magnetic fields can be varied to distinguish the $P,T$-violating transition amplitude from that due to the magnetic field.}
% The Rabi frequency changes by $2 \Omega_{PT}$ between $\beta=0$ and $\beta=\pi$. }
\label{fig:rabi_oscillations}
\end{figure}

We assume that the electric and magnetic fields are driven at the same frequency, $\omega_E = \omega_B = \omega$,
%We denote the energy separation between the clock states %in the presence of $\Bsca_z(t), \Esca_z(t)$ %(\emph{i.e.}, including the zero-field hyperfine splitting, and the Zeeman and tensor Stark shifts) 
%as $\omega_0$, 
with detuning $\Delta = \omega-\omega_0$. In the rotating wave approximation, %for the time-dependent fields %(assuming that the Bloch-Siegert shift from the counter-rotating term is also included in $\omega_0$). Then 
the Hamiltonian from Eq.\ (\ref{eq:hint_2}) %, in the two-dimensional subspace spanned by $\ket{g}$ and $\ket{e}$, 
is
\begin{equation}
\begin{split}
H_\mrm{eff} & = \frac{\Omega_B}{2} \, \sigma_x + \frac{\Omega_{PT}}{2} \left( \cos\beta \, \sigma_x + \sin\beta \, \sigma_y \right) + \frac{\Delta}{2}\sigma_z,
\end{split}
\end{equation}
where $\sigma_{x,y,z}$ are Pauli matrices, and the Rabi frequencies for the Zeeman and $P,T$-violating interactions are respectively $\Omega_B = -\frac{1}{2} (g_S - g_I)\mu_B \Bsca_0$ and $\Omega_{PT} = \frac{1}{2}W_{PT} \zeta_0$. When the molecule is driven on resonance ($\Delta=0$) for a time $\tau$, molecules initially prepared in $\ket{g}$ are transferred to $\ket{e}$. The excited-state population is then $\rho_{ee}(\tau) = \sin^2\left[\frac{(\Omega_B + \Omega_{PT} \, \cos \beta)\,\tau}{2} \right]$, where the vanishingly small terms that are quadratic in $W_{PT}$ have been dropped. The time evolution of $\rho_{ee}(\tau)$ and the effect of the phase $\beta$ are shown in Fig.\ \ref{fig:rabi_oscillations}.

For an EDM measurement, the magnetic field amplitude, $\Bsca_0$, and the pulse duration, $\tau$, are set so that $\Omega_B \tau \approx \pm \frac{\pi}{2}$ mod $2 \pi$, to make $\rho_{ee}$ maximally sensitive to $\Omega_{PT}$. The measurement is then repeated for different values of the phase angle $\beta$, as shown in Fig.\ \ref{fig:rabi_oscillations}. Setting $\beta= \pm \frac{\pi}{2}$ leaves $\rho_{ee}$ unchanged, and provides a convenient null test. The values of $\Bsca_0$ and $\tau$ can also be varied over a large dynamic range while maintaining the condition $\Omega_B \tau = \pm \frac{\pi}{2} \ \textrm{mod} \ 2 \pi$, which is a useful way to tease out systematic errors. 

A genuine $P,T$-violating signal can be identified as the part of $\rho_{ee}$ that changes sign under switches of (i) the initial state between $\ket{g}$ and $\ket{e}$, (ii) the phase $\beta$ between 0 and $\pi$, and (iii) the pulse area $\Omega_B \tau$ between $\pm \frac{\pi}{2}$ mod $2\pi$. 

%The $P,T$-violating signal of interest is contained in the rate of population transfer, the Rabi frequency, between the ground and excited states.  Where $W_{PT}$ increases (decreases) the Rabi freuqency for $\beta < \pi/2$ ($\beta > \pi/2$), with a maximum Rabi frequency when the oscillating electric field that polarizes the molecule and the Zeeman field that drives the clock transition are in-phase, \textit{i.e.} at $\beta = 0$, (and a corresponding minimum Rabi freuqency when $\beta = \pi$). 

%Changing the phase $\beta$ from 0 to $\pi$ in a traditional EDM measurement is analogous to changing the sign of the DC polarizaing electric field, where the large electric field reversal is often a source of troublesome systematics \cite{Graner2016}.  Phases of radio-frequency fields, on the other hand, can be changed smoothly. This provides the additional benefit of a continuous range of values (from 0 to $2\pi$) compared to only two discrete electric field values.

With measurements on a total of $N_\mathrm{tot}$ molecules, and an interaction time $\tau$ for each measurement cycle, the precision achievable in a projection-noise-limited measurement of $W_{PT} = 2 \Omega_{PT}/\zeta_0$ is $\delta W_{PT} = \frac{2}{\zeta_0 \tau \sqrt{N_\mathrm{tot}}}$. The  corresponding electron EDM precision (assuming that the electron EDM is the only source of $P,T$-violation) is $\delta d_e = \delta W_{PT}/2 e \Esca_\mrm{eff}$, where $\Esca_\mrm{eff}$ is the effective electric field experienced by the electron EDM in the molecule \cite{DeMille2017,Safronova2018}. For an experiment using our method with YbAg molecules, we estimate an electron EDM sensitivity 
\begin{widetext}
\begin{equation}
\delta d_e = 10^{-31} \ e \ \mrm{cm} \left( \frac{10^4}{N} \right)^{1/2} \left(\frac{10 \ \mrm{s}}{\tau}\right) \left(\frac{10 \ \mrm{d}}{T}\right)^{1/2}  \left(\frac{20 \ \mrm{GV/cm}}{\Esca_\mrm{eff}} \right)  \left( \frac{1}{\zeta_0} \right).
\end{equation}
\end{widetext}
Here $N$ is the number of  trapped molecules used per measurement cycle, and $T$ is the total integration time of the experiment. %This corresponds to a measurement of the $P,T$-violating Rabi frequency $\Omega_{PT}$ with a precision of $2\pi \times 0.5$ $\mu$Hz. 
We have assumed $\Esca_\mrm{eff} \sim$ 20 GV/cm in YbAg, similar to the value for the closely related YbF molecule \cite{Fleig,Abe2014}. An electron EDM measurement with a precision of $10^{-31} \ e \ \mrm{cm}$ would improve on the current state of the art by two orders of magnitude, and probe energy scales well beyond 100 TeV \cite{DeMille2017,Safronova2018}.

%%%%%%%%%%%%%%%%%%%%%%%%
\textit{Advantages. --} Compared to the traditional EDM search methods, a number of practical advantages are enabled by the clock state technique.
\begin{enumerate}[label=\alph*)]
    \item In the traditional method spurious low-frequency magnetic fields, \emph{e.g.} from leakage currents, are a common source of systematic errors. However in our method, only radio-frequency (rf) magnetic fields, in the spectral range $\omega_0 \pm 1/\tau$, can cause shifts of the Rabi frequency. Such magnetic fields are significantly easier to measure, and shield in practice (due to the high effectiveness of eddy current shielding at these frequencies), with consequent improvements to the control of systematic errors.  % It is also much easier to identify the source of an rf field compared to some slow DC source.

    \item The phase $\beta$ between the electric and magnetic fields can be smoothly varied with high precision using standard rf instruments. This eliminates switching transients and discharges, and the resulting magnetic field errors, that are encountered in typical EDM experiments where the sign of DC electric fields has to be switched. A number of potential systematic effects (discussed further below) also have a characteristic dependence on $\beta$ that is different from the $\cos \beta$ dependence of the $P,T$-violating interaction, which thus allows them to be cleanly separated from genuine new physics signals.  

    \item The hyperfine levels separated by $\omega_0$ are insensitive to magnetic field noise and fluctuations to first order. So the requirements for shielding stray magnetic fields and their low-frequency drifts are significantly relaxed compared to traditional EDM search experiments. This feature can lead to better control over systematics and simplify the design of experiments. 
    
    \item The magnetic field insensitivity of the clock states also improves the precision of EDM measurements, since long coherence times for hyperfine state superpositions can be realized \cite{Park2017}. 
    
    \item The initial state preparation is simple: it is easy to accurately initialize molecules in one of the hyperfine states $\ket{g}$ or $\ket{e}$, compared to preparing the spin-state superpositions used in the traditional EDM measurement method (e.g., \cite{Hudson2011}). This feature improves the duty cycle of experiments leading to better precision in a given integration time, and greatly reduces systematic errors due to imperfect state preparation.
            
    \item While the above analysis focuses on electron EDM measurements for the sake of simplicity, we note that our method can also be applied to nuclear EDM measurements in molecules with hyperfine clock states. This feature can be an advantage for nuclear EDM searches using radioactive nuclei that are extremely sensitive to $P,T$-violation (e.g., $^{225}$Ra \cite{Bishof2016} or $^{229}$Pa \cite{Singh2019}), but where there may only be a limited range of molecular species that can be efficiently produced from rare isotope sources. Our method can also be easily applied to a large class of molecular ions (\cite{SM}, Sec.\ E), since ion traps can be engineered to make their secular and micromotion frequencies be well-separated from the hyperfine resonance.
\end{enumerate}

%%%%%%%%%%%%%%%%%%%%%%%%%%%%%%%%%%
\textit{Controlling systematic errors. -- } 
%In general, any transition amplitudes that are in phase with the driving electric field could potentially lead to systematic errors.
%For example, $\Esca$-correlated $\Bsca$-fields can be produced by leakage currents between the electrodes, which can then drive the regular $M1$ transition between the clock states in a way that mimics a $P,T$-violating signal. 
%One strategy for detecting such errors is to make measurements in states (\textit{e.g.} $m_F=0 \to m_F=0$ clock transitions in the $N=0$ and $N=1$ rotational manifolds) that have similar responses to magnetic fields but different magnitudes and signs of $\zeta$. Such states can be used as internal co-magnetometers even in simple $^2\Sigma$ molecules \cite{Prasannaa2015}2015}. However, a 
Stringent control of systematic errors is an extremely important aspect of EDM measurements. Here we consider potential EDM-mimicking effects that are specific to our measurement technique. One possible source of such errors is the displacement current produced by the oscillating $\Esca$-field, which produces an $\Esca$-linear magnetic field. These systematic errors can be suppressed by many orders of magnitude, using the characteristic dependence of the $P,T$-violating signal on the phase $\beta$ and frequency $\omega_E$ of the $\Esca$-field (\cite{SM}, Sec.\ C). 

We have also analyzed other sources of systematic errors, such as $E1-M1$ mixing and differential Stark shifts. Stray background electric ($\Esca_\mrm{dc}$) and magnetic ($\Bsca_\mrm{dc}$) fields can admix rotational states in the $N=0$ and $N=1$ manifolds. This leads to a transition Rabi frequency $\Omega_{E1-M1} \propto \Esca_\mrm{dc} \Bsca_\mrm{dc} \Esca_0$ which can mimic $\Omega_{PT}$. For realistic estimates of the stray fields, numerical and analytical calculations described in the Supplementary Material indicate that this effect leads to a negligible systematic error. We have also considered systematic errors that arise through the differential Stark shift (DSS) of the hyperfine clock states in the oscillating electric field. Using both analytical and numerical models, we find that the DSS-induced error is also negligible (\cite{SM}, Sec.\ D). 

Finally, we point out that the nonlinear response of the molecular orientation to the $\Esca$-field provides a unique and powerful method to control any residual systematic errors. This method employs the fact that $\zeta(t)$ contains odd harmonics of the electric field frequency $\omega_E$ (see \cite{SM}, Fig.\ \ref{fig:harmonics}). So the electric field frequency can be set to a sub-harmonic of the hyperfine resonance frequency $\omega_0$ (e.g., $\omega_E = \omega_0/3$), while still allowing the $P,T$-violating interaction to \emph{resonantly} drive the hyperfine transition. But any systematic effects linear in the electric field (a condition that encompasses the overwhelming majority of them) are pushed far off resonance, and so their interference with the transition amplitude is highly suppressed. Sub-harmonic modulation therefore offers a clear and general diagnostic to distinguish genuine $P,T$-violating effects from spurious backgrounds. Precision control of electric and magnetic field amplitudes and phases in the rf domain, in combination with methods such as sub-harmonic modulation, provides a versatile toolbox to control systematic errors in EDM measurements using hyperfine clock transitions.

\textit{Summary. --} We have presented a technique for measuring parity and time-reversal violation, which leverages the magnetic-field insensitivity of ubiquitously available hyperfine clock transitions in polar molecules. The use of clock transitions enables longer coherence times leading to improved precision, and opens up new ways to control systematic errors in experiments using trapped ultracold molecules. A wide selection of ultracold molecules, including simple $^2\Sigma$ diatomic molecules that can be assembled out of ultracold atoms, thus becomes available for new physics searches.
\\ ~ \\
% \textit{Acknowledgments. --} 
We thank Wes Campbell, Jonathan Weinstein and David DeMille for helpful comments. M.V. acknowledges support from NSERC and an Ontario Graduate Scholarship. A.M.J. acknowledges support from NSF Grant No. PHY-1912665. A.C.V. acknowledges support from Canada Research Chairs and a Sloan Fellowship.

\bibliography{hfs_edm}

%merlin.mbs apsrev4-1.bst 2010-07-25 4.21a (PWD, AO, DPC) hacked
%Control: key (0)
%Control: author (8) initials jnrlst
%Control: editor formatted (1) identically to author
%Control: production of article title (-1) disabled
%Control: page (0) single
%Control: year (1) truncated
%Control: production of eprint (0) enabled
\begin{thebibliography}{42}%
\makeatletter
\providecommand \@ifxundefined [1]{%
 \@ifx{#1\undefined}
}%
\providecommand \@ifnum [1]{%
 \ifnum #1\expandafter \@firstoftwo
 \else \expandafter \@secondoftwo
 \fi
}%
\providecommand \@ifx [1]{%
 \ifx #1\expandafter \@firstoftwo
 \else \expandafter \@secondoftwo
 \fi
}%
\providecommand \natexlab [1]{#1}%
\providecommand \enquote  [1]{``#1''}%
\providecommand \bibnamefont  [1]{#1}%
\providecommand \bibfnamefont [1]{#1}%
\providecommand \citenamefont [1]{#1}%
\providecommand \href@noop [0]{\@secondoftwo}%
\providecommand \href [0]{\begingroup \@sanitize@url \@href}%
\providecommand \@href[1]{\@@startlink{#1}\@@href}%
\providecommand \@@href[1]{\endgroup#1\@@endlink}%
\providecommand \@sanitize@url [0]{\catcode `\\12\catcode `\$12\catcode
  `\&12\catcode `\#12\catcode `\^12\catcode `\_12\catcode `\%12\relax}%
\providecommand \@@startlink[1]{}%
\providecommand \@@endlink[0]{}%
\providecommand \url  [0]{\begingroup\@sanitize@url \@url }%
\providecommand \@url [1]{\endgroup\@href {#1}{\urlprefix }}%
\providecommand \urlprefix  [0]{URL }%
\providecommand \Eprint [0]{\href }%
\providecommand \doibase [0]{http://dx.doi.org/}%
\providecommand \selectlanguage [0]{\@gobble}%
\providecommand \bibinfo  [0]{\@secondoftwo}%
\providecommand \bibfield  [0]{\@secondoftwo}%
\providecommand \translation [1]{[#1]}%
\providecommand \BibitemOpen [0]{}%
\providecommand \bibitemStop [0]{}%
\providecommand \bibitemNoStop [0]{.\EOS\space}%
\providecommand \EOS [0]{\spacefactor3000\relax}%
\providecommand \BibitemShut  [1]{\csname bibitem#1\endcsname}%
\let\auto@bib@innerbib\@empty
%</preamble>
\bibitem [{\citenamefont {DeMille}\ \emph {et~al.}(2017)\citenamefont
  {DeMille}, \citenamefont {Doyle},\ and\ \citenamefont
  {Sushkov}}]{DeMille2017}%
  \BibitemOpen
  \bibfield  {author} {\bibinfo {author} {\bibfnamefont {D.}~\bibnamefont
  {DeMille}}, \bibinfo {author} {\bibfnamefont {J.~M.}\ \bibnamefont {Doyle}},
  \ and\ \bibinfo {author} {\bibfnamefont {A.~O.}\ \bibnamefont {Sushkov}},\
  }\href {\doibase 10.1126/science.aal3003} {\bibfield  {journal} {\bibinfo
  {journal} {Science}\ }\textbf {\bibinfo {volume} {357}},\ \bibinfo {pages}
  {990} (\bibinfo {year} {2017})}\BibitemShut {NoStop}%
\bibitem [{\citenamefont {Safronova}\ \emph {et~al.}(2018)\citenamefont
  {Safronova}, \citenamefont {Budker}, \citenamefont {DeMille}, \citenamefont
  {Kimball}, \citenamefont {Derevianko},\ and\ \citenamefont
  {Clark}}]{Safronova2018}%
  \BibitemOpen
  \bibfield  {author} {\bibinfo {author} {\bibfnamefont {M.~S.}\ \bibnamefont
  {Safronova}}, \bibinfo {author} {\bibfnamefont {D.}~\bibnamefont {Budker}},
  \bibinfo {author} {\bibfnamefont {D.}~\bibnamefont {DeMille}}, \bibinfo
  {author} {\bibfnamefont {D.~F.~J.}\ \bibnamefont {Kimball}}, \bibinfo
  {author} {\bibfnamefont {A.}~\bibnamefont {Derevianko}}, \ and\ \bibinfo
  {author} {\bibfnamefont {C.~W.}\ \bibnamefont {Clark}},\ }\href
  {https://link.aps.org/doi/10.1103/RevModPhys.90.025008} {\bibfield  {journal}
  {\bibinfo  {journal} {Rev. Mod. Phys.}\ }\textbf {\bibinfo {volume} {90}},\
  \bibinfo {pages} {025008} (\bibinfo {year} {2018})}\BibitemShut {NoStop}%
\bibitem [{\citenamefont {Hudson}\ \emph {et~al.}(2011)\citenamefont {Hudson},
  \citenamefont {Kara}, \citenamefont {Smallman}, \citenamefont {Sauer},
  \citenamefont {Tarbutt},\ and\ \citenamefont {Hinds}}]{Hudson2011}%
  \BibitemOpen
  \bibfield  {author} {\bibinfo {author} {\bibfnamefont {J.~J.}\ \bibnamefont
  {Hudson}}, \bibinfo {author} {\bibfnamefont {D.~M.}\ \bibnamefont {Kara}},
  \bibinfo {author} {\bibfnamefont {I.}~\bibnamefont {Smallman}}, \bibinfo
  {author} {\bibfnamefont {B.~E.}\ \bibnamefont {Sauer}}, \bibinfo {author}
  {\bibfnamefont {M.~R.}\ \bibnamefont {Tarbutt}}, \ and\ \bibinfo {author}
  {\bibfnamefont {E.~A.}\ \bibnamefont {Hinds}},\ }\href@noop {} {\bibfield
  {journal} {\bibinfo  {journal} {Nature}\ }\textbf {\bibinfo {volume} {473}},\
  \bibinfo {pages} {493} (\bibinfo {year} {2011})}\BibitemShut {NoStop}%
\bibitem [{\citenamefont {{ACME Collaboration: J. Baron}}\ \emph
  {et~al.}(2014)\citenamefont {{ACME Collaboration: J. Baron}} \emph
  {et~al.}}]{Baron2014}%
  \BibitemOpen
  \bibfield  {author} {\bibinfo {author} {\bibnamefont {{ACME Collaboration: J.
  Baron}}} \emph {et~al.},\ }\href@noop {} {\bibfield  {journal} {\bibinfo
  {journal} {Science}\ }\textbf {\bibinfo {volume} {343}},\ \bibinfo {pages}
  {269} (\bibinfo {year} {2014})}\BibitemShut {NoStop}%
\bibitem [{\citenamefont {Cairncross}\ \emph {et~al.}(2017)\citenamefont
  {Cairncross}, \citenamefont {Gresh}, \citenamefont {Grau}, \citenamefont
  {Cossel}, \citenamefont {Roussy}, \citenamefont {Ni}, \citenamefont {Zhou},
  \citenamefont {Ye},\ and\ \citenamefont {Cornell}}]{Cairncross2017}%
  \BibitemOpen
  \bibfield  {author} {\bibinfo {author} {\bibfnamefont {W.~B.}\ \bibnamefont
  {Cairncross}}, \bibinfo {author} {\bibfnamefont {D.~N.}\ \bibnamefont
  {Gresh}}, \bibinfo {author} {\bibfnamefont {M.}~\bibnamefont {Grau}},
  \bibinfo {author} {\bibfnamefont {K.~C.}\ \bibnamefont {Cossel}}, \bibinfo
  {author} {\bibfnamefont {T.~S.}\ \bibnamefont {Roussy}}, \bibinfo {author}
  {\bibfnamefont {Y.}~\bibnamefont {Ni}}, \bibinfo {author} {\bibfnamefont
  {Y.}~\bibnamefont {Zhou}}, \bibinfo {author} {\bibfnamefont {J.}~\bibnamefont
  {Ye}}, \ and\ \bibinfo {author} {\bibfnamefont {E.~A.}\ \bibnamefont
  {Cornell}},\ }\href {\doibase 10.1103/PhysRevLett.119.153001} {\bibfield
  {journal} {\bibinfo  {journal} {Phys. Rev. Lett.}\ }\textbf {\bibinfo
  {volume} {119}},\ \bibinfo {pages} {153001} (\bibinfo {year}
  {2017})}\BibitemShut {NoStop}%
\bibitem [{\citenamefont {{ACME Collaboration: V. Andreev}}\ \emph
  {et~al.}(2018)\citenamefont {{ACME Collaboration: V. Andreev}} \emph
  {et~al.}}]{Andreev2018}%
  \BibitemOpen
  \bibfield  {author} {\bibinfo {author} {\bibnamefont {{ACME Collaboration: V.
  Andreev}}} \emph {et~al.},\ }\href {\doibase 10.1038/s41586-018-0599-8}
  {\bibfield  {journal} {\bibinfo  {journal} {Nature}\ }\textbf {\bibinfo
  {volume} {562}},\ \bibinfo {pages} {355} (\bibinfo {year}
  {2018})}\BibitemShut {NoStop}%
\bibitem [{\citenamefont {Feng}(2013)}]{Feng2013}%
  \BibitemOpen
  \bibfield  {author} {\bibinfo {author} {\bibfnamefont {J.~L.}\ \bibnamefont
  {Feng}},\ }\href@noop {} {\bibfield  {journal} {\bibinfo  {journal} {Annu.
  Rev. Nucl. Part. Sci.}\ }\textbf {\bibinfo {volume} {63}},\ \bibinfo {pages}
  {351} (\bibinfo {year} {2013})}\BibitemShut {NoStop}%
\bibitem [{\citenamefont {Julienne}(2018)}]{Julienne2018}%
  \BibitemOpen
  \bibfield  {author} {\bibinfo {author} {\bibfnamefont {P.~S.}\ \bibnamefont
  {Julienne}},\ }\href@noop {} {\bibfield  {journal} {\bibinfo  {journal}
  {Nature Physics}\ }\textbf {\bibinfo {volume} {14}},\ \bibinfo {pages} {873}
  (\bibinfo {year} {2018})}\BibitemShut {NoStop}%
\bibitem [{\citenamefont {Truppe}\ \emph {et~al.}(2017)\citenamefont {Truppe},
  \citenamefont {Williams}, \citenamefont {Hambach}, \citenamefont {Caldwell},
  \citenamefont {Fitch}, \citenamefont {Hinds}, \citenamefont {Sauer},\ and\
  \citenamefont {Tarbutt}}]{Truppe2017}%
  \BibitemOpen
  \bibfield  {author} {\bibinfo {author} {\bibfnamefont {S.}~\bibnamefont
  {Truppe}}, \bibinfo {author} {\bibfnamefont {H.}~\bibnamefont {Williams}},
  \bibinfo {author} {\bibfnamefont {M.}~\bibnamefont {Hambach}}, \bibinfo
  {author} {\bibfnamefont {L.}~\bibnamefont {Caldwell}}, \bibinfo {author}
  {\bibfnamefont {N.}~\bibnamefont {Fitch}}, \bibinfo {author} {\bibfnamefont
  {E.}~\bibnamefont {Hinds}}, \bibinfo {author} {\bibfnamefont
  {B.}~\bibnamefont {Sauer}}, \ and\ \bibinfo {author} {\bibfnamefont
  {M.}~\bibnamefont {Tarbutt}},\ }\href@noop {} {\bibfield  {journal} {\bibinfo
   {journal} {Nat. Phys.}\ }\textbf {\bibinfo {volume} {13}},\ \bibinfo {pages}
  {1173} (\bibinfo {year} {2017})}\BibitemShut {NoStop}%
\bibitem [{\citenamefont {Collopy}\ \emph {et~al.}(2018)\citenamefont
  {Collopy}, \citenamefont {Ding}, \citenamefont {Wu}, \citenamefont
  {Finneran}, \citenamefont {Anderegg}, \citenamefont {Augenbraun},
  \citenamefont {Doyle},\ and\ \citenamefont {Ye}}]{Collopy2018}%
  \BibitemOpen
  \bibfield  {author} {\bibinfo {author} {\bibfnamefont {A.~L.}\ \bibnamefont
  {Collopy}}, \bibinfo {author} {\bibfnamefont {S.}~\bibnamefont {Ding}},
  \bibinfo {author} {\bibfnamefont {Y.}~\bibnamefont {Wu}}, \bibinfo {author}
  {\bibfnamefont {I.~A.}\ \bibnamefont {Finneran}}, \bibinfo {author}
  {\bibfnamefont {L.}~\bibnamefont {Anderegg}}, \bibinfo {author}
  {\bibfnamefont {B.~L.}\ \bibnamefont {Augenbraun}}, \bibinfo {author}
  {\bibfnamefont {J.~M.}\ \bibnamefont {Doyle}}, \ and\ \bibinfo {author}
  {\bibfnamefont {J.}~\bibnamefont {Ye}},\ }\href {\doibase
  10.1103/PhysRevLett.121.213201} {\bibfield  {journal} {\bibinfo  {journal}
  {Phys. Rev. Lett.}\ }\textbf {\bibinfo {volume} {121}},\ \bibinfo {pages}
  {213201} (\bibinfo {year} {2018})}\BibitemShut {NoStop}%
\bibitem [{\citenamefont {Cheuk}\ \emph {et~al.}(2018)\citenamefont {Cheuk},
  \citenamefont {Anderegg}, \citenamefont {Augenbraun}, \citenamefont {Bao},
  \citenamefont {Burchesky}, \citenamefont {Ketterle},\ and\ \citenamefont
  {Doyle}}]{Cheuk2018}%
  \BibitemOpen
  \bibfield  {author} {\bibinfo {author} {\bibfnamefont {L.~W.}\ \bibnamefont
  {Cheuk}}, \bibinfo {author} {\bibfnamefont {L.}~\bibnamefont {Anderegg}},
  \bibinfo {author} {\bibfnamefont {B.~L.}\ \bibnamefont {Augenbraun}},
  \bibinfo {author} {\bibfnamefont {Y.}~\bibnamefont {Bao}}, \bibinfo {author}
  {\bibfnamefont {S.}~\bibnamefont {Burchesky}}, \bibinfo {author}
  {\bibfnamefont {W.}~\bibnamefont {Ketterle}}, \ and\ \bibinfo {author}
  {\bibfnamefont {J.~M.}\ \bibnamefont {Doyle}},\ }\href {\doibase
  10.1103/PhysRevLett.121.083201} {\bibfield  {journal} {\bibinfo  {journal}
  {Phys. Rev. Lett.}\ }\textbf {\bibinfo {volume} {121}},\ \bibinfo {pages}
  {083201} (\bibinfo {year} {2018})}\BibitemShut {NoStop}%
\bibitem [{\citenamefont {Molony}\ \emph {et~al.}(2014)\citenamefont {Molony},
  \citenamefont {Gregory}, \citenamefont {Ji}, \citenamefont {Lu},
  \citenamefont {K\"oppinger}, \citenamefont {Le~Sueur}, \citenamefont
  {Blackley}, \citenamefont {Hutson},\ and\ \citenamefont
  {Cornish}}]{Molony2014}%
  \BibitemOpen
  \bibfield  {author} {\bibinfo {author} {\bibfnamefont {P.~K.}\ \bibnamefont
  {Molony}}, \bibinfo {author} {\bibfnamefont {P.~D.}\ \bibnamefont {Gregory}},
  \bibinfo {author} {\bibfnamefont {Z.}~\bibnamefont {Ji}}, \bibinfo {author}
  {\bibfnamefont {B.}~\bibnamefont {Lu}}, \bibinfo {author} {\bibfnamefont
  {M.~P.}\ \bibnamefont {K\"oppinger}}, \bibinfo {author} {\bibfnamefont
  {C.~R.}\ \bibnamefont {Le~Sueur}}, \bibinfo {author} {\bibfnamefont {C.~L.}\
  \bibnamefont {Blackley}}, \bibinfo {author} {\bibfnamefont {J.~M.}\
  \bibnamefont {Hutson}}, \ and\ \bibinfo {author} {\bibfnamefont {S.~L.}\
  \bibnamefont {Cornish}},\ }\href {\doibase 10.1103/PhysRevLett.113.255301}
  {\bibfield  {journal} {\bibinfo  {journal} {Phys. Rev. Lett.}\ }\textbf
  {\bibinfo {volume} {113}},\ \bibinfo {pages} {255301} (\bibinfo {year}
  {2014})}\BibitemShut {NoStop}%
\bibitem [{\citenamefont {Takekoshi}\ \emph {et~al.}(2014)\citenamefont
  {Takekoshi}, \citenamefont {Reichs\"ollner}, \citenamefont {Schindewolf},
  \citenamefont {Hutson}, \citenamefont {Le~Sueur}, \citenamefont {Dulieu},
  \citenamefont {Ferlaino}, \citenamefont {Grimm},\ and\ \citenamefont
  {N\"agerl}}]{Takekoshi2014}%
  \BibitemOpen
  \bibfield  {author} {\bibinfo {author} {\bibfnamefont {T.}~\bibnamefont
  {Takekoshi}}, \bibinfo {author} {\bibfnamefont {L.}~\bibnamefont
  {Reichs\"ollner}}, \bibinfo {author} {\bibfnamefont {A.}~\bibnamefont
  {Schindewolf}}, \bibinfo {author} {\bibfnamefont {J.~M.}\ \bibnamefont
  {Hutson}}, \bibinfo {author} {\bibfnamefont {C.~R.}\ \bibnamefont
  {Le~Sueur}}, \bibinfo {author} {\bibfnamefont {O.}~\bibnamefont {Dulieu}},
  \bibinfo {author} {\bibfnamefont {F.}~\bibnamefont {Ferlaino}}, \bibinfo
  {author} {\bibfnamefont {R.}~\bibnamefont {Grimm}}, \ and\ \bibinfo {author}
  {\bibfnamefont {H.-C.}\ \bibnamefont {N\"agerl}},\ }\href {\doibase
  10.1103/PhysRevLett.113.205301} {\bibfield  {journal} {\bibinfo  {journal}
  {Phys. Rev. Lett.}\ }\textbf {\bibinfo {volume} {113}},\ \bibinfo {pages}
  {205301} (\bibinfo {year} {2014})}\BibitemShut {NoStop}%
\bibitem [{\citenamefont {Park}\ \emph {et~al.}(2015)\citenamefont {Park},
  \citenamefont {Will},\ and\ \citenamefont {Zwierlein}}]{Park2015}%
  \BibitemOpen
  \bibfield  {author} {\bibinfo {author} {\bibfnamefont {J.~W.}\ \bibnamefont
  {Park}}, \bibinfo {author} {\bibfnamefont {S.~A.}\ \bibnamefont {Will}}, \
  and\ \bibinfo {author} {\bibfnamefont {M.~W.}\ \bibnamefont {Zwierlein}},\
  }\href {\doibase 10.1103/PhysRevLett.114.205302} {\bibfield  {journal}
  {\bibinfo  {journal} {Phys. Rev. Lett.}\ }\textbf {\bibinfo {volume} {114}},\
  \bibinfo {pages} {205302} (\bibinfo {year} {2015})}\BibitemShut {NoStop}%
\bibitem [{\citenamefont {Guo}\ \emph {et~al.}(2016)\citenamefont {Guo},
  \citenamefont {Zhu}, \citenamefont {Lu}, \citenamefont {Ye}, \citenamefont
  {Wang}, \citenamefont {Vexiau}, \citenamefont {Bouloufa-Maafa}, \citenamefont
  {Qu\'em\'ener}, \citenamefont {Dulieu},\ and\ \citenamefont
  {Wang}}]{Guo2016}%
  \BibitemOpen
  \bibfield  {author} {\bibinfo {author} {\bibfnamefont {M.}~\bibnamefont
  {Guo}}, \bibinfo {author} {\bibfnamefont {B.}~\bibnamefont {Zhu}}, \bibinfo
  {author} {\bibfnamefont {B.}~\bibnamefont {Lu}}, \bibinfo {author}
  {\bibfnamefont {X.}~\bibnamefont {Ye}}, \bibinfo {author} {\bibfnamefont
  {F.}~\bibnamefont {Wang}}, \bibinfo {author} {\bibfnamefont {R.}~\bibnamefont
  {Vexiau}}, \bibinfo {author} {\bibfnamefont {N.}~\bibnamefont
  {Bouloufa-Maafa}}, \bibinfo {author} {\bibfnamefont {G.}~\bibnamefont
  {Qu\'em\'ener}}, \bibinfo {author} {\bibfnamefont {O.}~\bibnamefont
  {Dulieu}}, \ and\ \bibinfo {author} {\bibfnamefont {D.}~\bibnamefont
  {Wang}},\ }\href {\doibase 10.1103/PhysRevLett.116.205303} {\bibfield
  {journal} {\bibinfo  {journal} {Phys. Rev. Lett.}\ }\textbf {\bibinfo
  {volume} {116}},\ \bibinfo {pages} {205303} (\bibinfo {year}
  {2016})}\BibitemShut {NoStop}%
\bibitem [{\citenamefont {Rvachov}\ \emph {et~al.}(2017)\citenamefont
  {Rvachov}, \citenamefont {Son}, \citenamefont {Sommer}, \citenamefont
  {Ebadi}, \citenamefont {Park}, \citenamefont {Zwierlein}, \citenamefont
  {Ketterle},\ and\ \citenamefont {Jamison}}]{Rvachov2017}%
  \BibitemOpen
  \bibfield  {author} {\bibinfo {author} {\bibfnamefont {T.~M.}\ \bibnamefont
  {Rvachov}}, \bibinfo {author} {\bibfnamefont {H.}~\bibnamefont {Son}},
  \bibinfo {author} {\bibfnamefont {A.~T.}\ \bibnamefont {Sommer}}, \bibinfo
  {author} {\bibfnamefont {S.}~\bibnamefont {Ebadi}}, \bibinfo {author}
  {\bibfnamefont {J.~J.}\ \bibnamefont {Park}}, \bibinfo {author}
  {\bibfnamefont {M.~W.}\ \bibnamefont {Zwierlein}}, \bibinfo {author}
  {\bibfnamefont {W.}~\bibnamefont {Ketterle}}, \ and\ \bibinfo {author}
  {\bibfnamefont {A.~O.}\ \bibnamefont {Jamison}},\ }\href@noop {} {\bibfield
  {journal} {\bibinfo  {journal} {Phys. Rev. Lett.}\ }\textbf {\bibinfo
  {volume} {119}},\ \bibinfo {pages} {143001} (\bibinfo {year}
  {2017})}\BibitemShut {NoStop}%
\bibitem [{\citenamefont {Barb{\'e}}\ \emph {et~al.}(2018)\citenamefont
  {Barb{\'e}}, \citenamefont {Ciamei}, \citenamefont {Pasquiou}, \citenamefont
  {Reichs{\"o}llner}, \citenamefont {Schreck}, \citenamefont {{\.Z}uchowski},\
  and\ \citenamefont {Hutson}}]{Barbe2018}%
  \BibitemOpen
  \bibfield  {author} {\bibinfo {author} {\bibfnamefont {V.}~\bibnamefont
  {Barb{\'e}}}, \bibinfo {author} {\bibfnamefont {A.}~\bibnamefont {Ciamei}},
  \bibinfo {author} {\bibfnamefont {B.}~\bibnamefont {Pasquiou}}, \bibinfo
  {author} {\bibfnamefont {L.}~\bibnamefont {Reichs{\"o}llner}}, \bibinfo
  {author} {\bibfnamefont {F.}~\bibnamefont {Schreck}}, \bibinfo {author}
  {\bibfnamefont {P.~S.}\ \bibnamefont {{\.Z}uchowski}}, \ and\ \bibinfo
  {author} {\bibfnamefont {J.~M.}\ \bibnamefont {Hutson}},\ }\href@noop {}
  {\bibfield  {journal} {\bibinfo  {journal} {Nature Phys.}\ }\textbf {\bibinfo
  {volume} {14}},\ \bibinfo {pages} {881} (\bibinfo {year} {2018})}\BibitemShut
  {NoStop}%
\bibitem [{\citenamefont {Guttridge}\ \emph {et~al.}(2018)\citenamefont
  {Guttridge}, \citenamefont {Hopkins}, \citenamefont {Frye}, \citenamefont
  {McFerran}, \citenamefont {Hutson},\ and\ \citenamefont
  {Cornish}}]{Guttridge2018}%
  \BibitemOpen
  \bibfield  {author} {\bibinfo {author} {\bibfnamefont {A.}~\bibnamefont
  {Guttridge}}, \bibinfo {author} {\bibfnamefont {S.~A.}\ \bibnamefont
  {Hopkins}}, \bibinfo {author} {\bibfnamefont {M.~D.}\ \bibnamefont {Frye}},
  \bibinfo {author} {\bibfnamefont {J.~J.}\ \bibnamefont {McFerran}}, \bibinfo
  {author} {\bibfnamefont {J.~M.}\ \bibnamefont {Hutson}}, \ and\ \bibinfo
  {author} {\bibfnamefont {S.~L.}\ \bibnamefont {Cornish}},\ }\href {\doibase
  10.1103/PhysRevA.97.063414} {\bibfield  {journal} {\bibinfo  {journal} {Phys.
  Rev. A}\ }\textbf {\bibinfo {volume} {97}},\ \bibinfo {pages} {063414}
  (\bibinfo {year} {2018})}\BibitemShut {NoStop}%
\bibitem [{\citenamefont {Green}\ \emph {et~al.}(2019)\citenamefont {Green},
  \citenamefont {See~Toh}, \citenamefont {Roy}, \citenamefont {Li},
  \citenamefont {Kotochigova},\ and\ \citenamefont {Gupta}}]{Green2019}%
  \BibitemOpen
  \bibfield  {author} {\bibinfo {author} {\bibfnamefont {A.}~\bibnamefont
  {Green}}, \bibinfo {author} {\bibfnamefont {J.~H.}\ \bibnamefont {See~Toh}},
  \bibinfo {author} {\bibfnamefont {R.}~\bibnamefont {Roy}}, \bibinfo {author}
  {\bibfnamefont {M.}~\bibnamefont {Li}}, \bibinfo {author} {\bibfnamefont
  {S.}~\bibnamefont {Kotochigova}}, \ and\ \bibinfo {author} {\bibfnamefont
  {S.}~\bibnamefont {Gupta}},\ }\href
  {https://link.aps.org/doi/10.1103/PhysRevA.99.063416} {\bibfield  {journal}
  {\bibinfo  {journal} {PRA}\ }\textbf {\bibinfo {volume} {99}},\ \bibinfo
  {pages} {063416} (\bibinfo {year} {2019})}\BibitemShut {NoStop}%
\bibitem [{\citenamefont {Hu}\ \emph {et~al.}(2019)\citenamefont {Hu},
  \citenamefont {Liu}, \citenamefont {Grimes}, \citenamefont {Lin},
  \citenamefont {Gheorghe}, \citenamefont {Vexiau}, \citenamefont
  {Bouloufa-Maafa}, \citenamefont {Dulieu}, \citenamefont {Rosenband},\ and\
  \citenamefont {Ni}}]{Hu2019}%
  \BibitemOpen
  \bibfield  {author} {\bibinfo {author} {\bibfnamefont {M.-G.}\ \bibnamefont
  {Hu}}, \bibinfo {author} {\bibfnamefont {Y.}~\bibnamefont {Liu}}, \bibinfo
  {author} {\bibfnamefont {D.}~\bibnamefont {Grimes}}, \bibinfo {author}
  {\bibfnamefont {Y.-W.}\ \bibnamefont {Lin}}, \bibinfo {author} {\bibfnamefont
  {A.}~\bibnamefont {Gheorghe}}, \bibinfo {author} {\bibfnamefont
  {R.}~\bibnamefont {Vexiau}}, \bibinfo {author} {\bibfnamefont
  {N.}~\bibnamefont {Bouloufa-Maafa}}, \bibinfo {author} {\bibfnamefont
  {O.}~\bibnamefont {Dulieu}}, \bibinfo {author} {\bibfnamefont
  {T.}~\bibnamefont {Rosenband}}, \ and\ \bibinfo {author} {\bibfnamefont
  {K.-K.}\ \bibnamefont {Ni}},\ }\href@noop {} {\bibfield  {journal} {\bibinfo
  {journal} {Science}\ }\textbf {\bibinfo {volume} {366}},\ \bibinfo {pages}
  {1111} (\bibinfo {year} {2019})}\BibitemShut {NoStop}%
\bibitem [{\citenamefont {De~Marco}\ \emph {et~al.}(2019)\citenamefont
  {De~Marco}, \citenamefont {Valtolina}, \citenamefont {Matsuda}, \citenamefont
  {Tobias}, \citenamefont {Covey},\ and\ \citenamefont {Ye}}]{DeMarco2019}%
  \BibitemOpen
  \bibfield  {author} {\bibinfo {author} {\bibfnamefont {L.}~\bibnamefont
  {De~Marco}}, \bibinfo {author} {\bibfnamefont {G.}~\bibnamefont {Valtolina}},
  \bibinfo {author} {\bibfnamefont {K.}~\bibnamefont {Matsuda}}, \bibinfo
  {author} {\bibfnamefont {W.~G.}\ \bibnamefont {Tobias}}, \bibinfo {author}
  {\bibfnamefont {J.~P.}\ \bibnamefont {Covey}}, \ and\ \bibinfo {author}
  {\bibfnamefont {J.}~\bibnamefont {Ye}},\ }\href
  {http://science.sciencemag.org/content/363/6429/853.abstract} {\bibfield
  {journal} {\bibinfo  {journal} {Science}\ }\textbf {\bibinfo {volume}
  {363}},\ \bibinfo {pages} {853} (\bibinfo {year} {2019})}\BibitemShut
  {NoStop}%
\bibitem [{\citenamefont {Sunaga}\ \emph {et~al.}(2019)\citenamefont {Sunaga},
  \citenamefont {Prasannaa}, \citenamefont {Abe}, \citenamefont {Hada},\ and\
  \citenamefont {Das}}]{Sunaga2018}%
  \BibitemOpen
  \bibfield  {author} {\bibinfo {author} {\bibfnamefont {A.}~\bibnamefont
  {Sunaga}}, \bibinfo {author} {\bibfnamefont {V.~S.}\ \bibnamefont
  {Prasannaa}}, \bibinfo {author} {\bibfnamefont {M.}~\bibnamefont {Abe}},
  \bibinfo {author} {\bibfnamefont {M.}~\bibnamefont {Hada}}, \ and\ \bibinfo
  {author} {\bibfnamefont {B.~P.}\ \bibnamefont {Das}},\ }\href {\doibase
  10.1103/PhysRevA.99.040501} {\bibfield  {journal} {\bibinfo  {journal} {Phys.
  Rev. A}\ }\textbf {\bibinfo {volume} {99}},\ \bibinfo {pages} {040501}
  (\bibinfo {year} {2019})}\BibitemShut {NoStop}%
\bibitem [{\citenamefont {Lim}\ \emph {et~al.}(2018)\citenamefont {Lim},
  \citenamefont {Almond}, \citenamefont {Trigatzis}, \citenamefont {Devlin},
  \citenamefont {Fitch}, \citenamefont {Sauer}, \citenamefont {Tarbutt},\ and\
  \citenamefont {Hinds}}]{Lim2018}%
  \BibitemOpen
  \bibfield  {author} {\bibinfo {author} {\bibfnamefont {J.}~\bibnamefont
  {Lim}}, \bibinfo {author} {\bibfnamefont {J.~R.}\ \bibnamefont {Almond}},
  \bibinfo {author} {\bibfnamefont {M.~A.}\ \bibnamefont {Trigatzis}}, \bibinfo
  {author} {\bibfnamefont {J.~A.}\ \bibnamefont {Devlin}}, \bibinfo {author}
  {\bibfnamefont {N.~J.}\ \bibnamefont {Fitch}}, \bibinfo {author}
  {\bibfnamefont {B.~E.}\ \bibnamefont {Sauer}}, \bibinfo {author}
  {\bibfnamefont {M.~R.}\ \bibnamefont {Tarbutt}}, \ and\ \bibinfo {author}
  {\bibfnamefont {E.~A.}\ \bibnamefont {Hinds}},\ }\href {\doibase
  10.1103/PhysRevLett.120.123201} {\bibfield  {journal} {\bibinfo  {journal}
  {Phys. Rev. Lett.}\ }\textbf {\bibinfo {volume} {120}},\ \bibinfo {pages}
  {123201} (\bibinfo {year} {2018})}\BibitemShut {NoStop}%
\bibitem [{\citenamefont {Fleig}\ and\ \citenamefont
  {DeMille}(2019)}]{Fleig_DeMille}%
  \BibitemOpen
  \bibfield  {author} {\bibinfo {author} {\bibfnamefont {T.}~\bibnamefont
  {Fleig}}\ and\ \bibinfo {author} {\bibfnamefont {D.}~\bibnamefont
  {DeMille}},\ }\href@noop {} {}\bibinfo {howpublished} {private communication}
  (\bibinfo {year} {2019})\BibitemShut {NoStop}%
\bibitem [{\citenamefont {Norrgard}\ \emph {et~al.}(2017)\citenamefont
  {Norrgard}, \citenamefont {Edwards}, \citenamefont {McCarron}, \citenamefont
  {Steinecker}, \citenamefont {DeMille}, \citenamefont {Alam}, \citenamefont
  {Peck}, \citenamefont {Wadia},\ and\ \citenamefont {Hunter}}]{Norrgard2017}%
  \BibitemOpen
  \bibfield  {author} {\bibinfo {author} {\bibfnamefont {E.~B.}\ \bibnamefont
  {Norrgard}}, \bibinfo {author} {\bibfnamefont {E.~R.}\ \bibnamefont
  {Edwards}}, \bibinfo {author} {\bibfnamefont {D.~J.}\ \bibnamefont
  {McCarron}}, \bibinfo {author} {\bibfnamefont {M.~H.}\ \bibnamefont
  {Steinecker}}, \bibinfo {author} {\bibfnamefont {D.}~\bibnamefont {DeMille}},
  \bibinfo {author} {\bibfnamefont {S.~S.}\ \bibnamefont {Alam}}, \bibinfo
  {author} {\bibfnamefont {S.~K.}\ \bibnamefont {Peck}}, \bibinfo {author}
  {\bibfnamefont {N.~S.}\ \bibnamefont {Wadia}}, \ and\ \bibinfo {author}
  {\bibfnamefont {L.~R.}\ \bibnamefont {Hunter}},\ }\href {\doibase
  10.1103/PhysRevA.95.062506} {\bibfield  {journal} {\bibinfo  {journal} {Phys.
  Rev. A}\ }\textbf {\bibinfo {volume} {95}},\ \bibinfo {pages} {062506}
  (\bibinfo {year} {2017})}\BibitemShut {NoStop}%
\bibitem [{\citenamefont {Kozyryev}\ and\ \citenamefont
  {Hutzler}(2017)}]{Kozyryev2017}%
  \BibitemOpen
  \bibfield  {author} {\bibinfo {author} {\bibfnamefont {I.}~\bibnamefont
  {Kozyryev}}\ and\ \bibinfo {author} {\bibfnamefont {N.~R.}\ \bibnamefont
  {Hutzler}},\ }\href@noop {} {\bibfield  {journal} {\bibinfo  {journal} {Phys.
  Rev. Lett.}\ }\textbf {\bibinfo {volume} {119}},\ \bibinfo {pages} {133002}
  (\bibinfo {year} {2017})}\BibitemShut {NoStop}%
\bibitem [{\citenamefont {Park}\ \emph {et~al.}(2017)\citenamefont {Park},
  \citenamefont {Yan}, \citenamefont {Loh}, \citenamefont {Will},\ and\
  \citenamefont {Zwierlein}}]{Park2017}%
  \BibitemOpen
  \bibfield  {author} {\bibinfo {author} {\bibfnamefont {J.~W.}\ \bibnamefont
  {Park}}, \bibinfo {author} {\bibfnamefont {Z.~Z.}\ \bibnamefont {Yan}},
  \bibinfo {author} {\bibfnamefont {H.}~\bibnamefont {Loh}}, \bibinfo {author}
  {\bibfnamefont {S.~A.}\ \bibnamefont {Will}}, \ and\ \bibinfo {author}
  {\bibfnamefont {M.~f.}\ \bibnamefont {Zwierlein}},\ }\href {\doibase
  10.1126/science.aal5066} {\bibfield  {journal} {\bibinfo  {journal}
  {Science}\ }\textbf {\bibinfo {volume} {357}},\ \bibinfo {pages} {372}
  (\bibinfo {year} {2017})}\BibitemShut {NoStop}%
\bibitem [{SM()}]{SM}%
  \BibitemOpen
  \href@noop {} {}\bibinfo {howpublished} {See Supplementary
  Material}\BibitemShut {NoStop}%
\bibitem [{\citenamefont {Meyer}\ and\ \citenamefont {Bohn}(2009)}]{Meyer2009}%
  \BibitemOpen
  \bibfield  {author} {\bibinfo {author} {\bibfnamefont {E.~R.}\ \bibnamefont
  {Meyer}}\ and\ \bibinfo {author} {\bibfnamefont {J.~L.}\ \bibnamefont
  {Bohn}},\ }\href {\doibase 10.1103/PhysRevA.80.042508} {\bibfield  {journal}
  {\bibinfo  {journal} {Phys. Rev. A}\ }\textbf {\bibinfo {volume} {80}},\
  \bibinfo {pages} {042508} (\bibinfo {year} {2009})}\BibitemShut {NoStop}%
\bibitem [{\citenamefont {Fleig}(2020)}]{Fleig}%
  \BibitemOpen
  \bibfield  {author} {\bibinfo {author} {\bibfnamefont {T.}~\bibnamefont
  {Fleig}},\ }\href@noop {} {}\bibinfo {howpublished} {private communication}
  (\bibinfo {year} {2020})\BibitemShut {NoStop}%
\bibitem [{\citenamefont {Abe}\ \emph {et~al.}(2014)\citenamefont {Abe},
  \citenamefont {Gopakumar}, \citenamefont {Hada}, \citenamefont {Das},
  \citenamefont {Tatewaki},\ and\ \citenamefont {Mukherjee}}]{Abe2014}%
  \BibitemOpen
  \bibfield  {author} {\bibinfo {author} {\bibfnamefont {M.}~\bibnamefont
  {Abe}}, \bibinfo {author} {\bibfnamefont {G.}~\bibnamefont {Gopakumar}},
  \bibinfo {author} {\bibfnamefont {M.}~\bibnamefont {Hada}}, \bibinfo {author}
  {\bibfnamefont {B.~P.}\ \bibnamefont {Das}}, \bibinfo {author} {\bibfnamefont
  {H.}~\bibnamefont {Tatewaki}}, \ and\ \bibinfo {author} {\bibfnamefont
  {D.}~\bibnamefont {Mukherjee}},\ }\href {\doibase 10.1103/PhysRevA.90.022501}
  {\bibfield  {journal} {\bibinfo  {journal} {Phys. Rev. A}\ }\textbf {\bibinfo
  {volume} {90}},\ \bibinfo {pages} {022501} (\bibinfo {year}
  {2014})}\BibitemShut {NoStop}%
\bibitem [{\citenamefont {Bishof}\ \emph {et~al.}(2016)\citenamefont {Bishof},
  \citenamefont {Parker}, \citenamefont {Bailey}, \citenamefont {Greene},
  \citenamefont {Holt}, \citenamefont {Kalita}, \citenamefont {Korsch},
  \citenamefont {Lemke}, \citenamefont {Lu}, \citenamefont {Mueller},
  \citenamefont {O'Connor}, \citenamefont {Singh},\ and\ \citenamefont
  {Dietrich}}]{Bishof2016}%
  \BibitemOpen
  \bibfield  {author} {\bibinfo {author} {\bibfnamefont {M.}~\bibnamefont
  {Bishof}}, \bibinfo {author} {\bibfnamefont {R.~H.}\ \bibnamefont {Parker}},
  \bibinfo {author} {\bibfnamefont {K.~G.}\ \bibnamefont {Bailey}}, \bibinfo
  {author} {\bibfnamefont {J.~P.}\ \bibnamefont {Greene}}, \bibinfo {author}
  {\bibfnamefont {R.~J.}\ \bibnamefont {Holt}}, \bibinfo {author}
  {\bibfnamefont {M.~R.}\ \bibnamefont {Kalita}}, \bibinfo {author}
  {\bibfnamefont {W.}~\bibnamefont {Korsch}}, \bibinfo {author} {\bibfnamefont
  {N.~D.}\ \bibnamefont {Lemke}}, \bibinfo {author} {\bibfnamefont {Z.-T.}\
  \bibnamefont {Lu}}, \bibinfo {author} {\bibfnamefont {P.}~\bibnamefont
  {Mueller}}, \bibinfo {author} {\bibfnamefont {T.~P.}\ \bibnamefont
  {O'Connor}}, \bibinfo {author} {\bibfnamefont {J.~T.}\ \bibnamefont {Singh}},
  \ and\ \bibinfo {author} {\bibfnamefont {M.~R.}\ \bibnamefont {Dietrich}},\
  }\href {\doibase 10.1103/PhysRevC.94.025501} {\bibfield  {journal} {\bibinfo
  {journal} {Phys. Rev. C}\ }\textbf {\bibinfo {volume} {94}},\ \bibinfo
  {pages} {025501} (\bibinfo {year} {2016})}\BibitemShut {NoStop}%
\bibitem [{\citenamefont {Singh}(2019)}]{Singh2019}%
  \BibitemOpen
  \bibfield  {author} {\bibinfo {author} {\bibfnamefont {J.~T.}\ \bibnamefont
  {Singh}},\ }\href {\doibase 10.1007/s10751-019-1573-z} {\bibfield  {journal}
  {\bibinfo  {journal} {Hyperfine Interactions}\ }\textbf {\bibinfo {volume}
  {240}},\ \bibinfo {pages} {29} (\bibinfo {year} {2019})}\BibitemShut
  {NoStop}%
\bibitem [{\citenamefont {Sauer}\ \emph {et~al.}(1996)\citenamefont {Sauer},
  \citenamefont {Wang},\ and\ \citenamefont {Hinds}}]{Sauer1996}%
  \BibitemOpen
  \bibfield  {author} {\bibinfo {author} {\bibfnamefont {B.}~\bibnamefont
  {Sauer}}, \bibinfo {author} {\bibfnamefont {J.}~\bibnamefont {Wang}}, \ and\
  \bibinfo {author} {\bibfnamefont {E.}~\bibnamefont {Hinds}},\ }\href
  {\doibase 10.1063/1.472569} {\bibfield  {journal} {\bibinfo  {journal} {J.
  Chem. Phys.}\ }\textbf {\bibinfo {volume} {105}},\ \bibinfo {pages} {7412}
  (\bibinfo {year} {1996})}\BibitemShut {NoStop}%
\bibitem [{\citenamefont {Khriplovich}\ and\ \citenamefont
  {Lamoreaux}(1997)}]{Khriplovich1997}%
  \BibitemOpen
  \bibfield  {author} {\bibinfo {author} {\bibfnamefont {I.~B.}\ \bibnamefont
  {Khriplovich}}\ and\ \bibinfo {author} {\bibfnamefont {S.~K.}\ \bibnamefont
  {Lamoreaux}},\ }\href@noop {} {\emph {\bibinfo {title} {CP violation without
  strangeness: electric dipole moments of particles, atoms, and molecules}}}\
  (\bibinfo  {publisher} {Springer-Verlag, Berlin},\ \bibinfo {year}
  {1997})\BibitemShut {NoStop}%
\bibitem [{\citenamefont {Hudson}\ \emph {et~al.}(2002)\citenamefont {Hudson},
  \citenamefont {Sauer}, \citenamefont {Tarbutt},\ and\ \citenamefont
  {Hinds}}]{Hudson2002}%
  \BibitemOpen
  \bibfield  {author} {\bibinfo {author} {\bibfnamefont {J.~J.}\ \bibnamefont
  {Hudson}}, \bibinfo {author} {\bibfnamefont {B.~E.}\ \bibnamefont {Sauer}},
  \bibinfo {author} {\bibfnamefont {M.~R.}\ \bibnamefont {Tarbutt}}, \ and\
  \bibinfo {author} {\bibfnamefont {E.~A.}\ \bibnamefont {Hinds}},\ }\href
  {\doibase 10.1103/PhysRevLett.89.023003} {\bibfield  {journal} {\bibinfo
  {journal} {Phys. Rev. Lett.}\ }\textbf {\bibinfo {volume} {89}},\ \bibinfo
  {pages} {023003} (\bibinfo {year} {2002})}\BibitemShut {NoStop}%
\bibitem [{\citenamefont {DeMille}\ \emph {et~al.}(2000)\citenamefont
  {DeMille}, \citenamefont {Bay}, \citenamefont {Bickman}, \citenamefont
  {Kawall}, \citenamefont {Krause}, \citenamefont {Maxwell},\ and\
  \citenamefont {Hunter}}]{DeMille2000}%
  \BibitemOpen
  \bibfield  {author} {\bibinfo {author} {\bibfnamefont {D.}~\bibnamefont
  {DeMille}}, \bibinfo {author} {\bibfnamefont {F.}~\bibnamefont {Bay}},
  \bibinfo {author} {\bibfnamefont {S.}~\bibnamefont {Bickman}}, \bibinfo
  {author} {\bibfnamefont {D.}~\bibnamefont {Kawall}}, \bibinfo {author}
  {\bibfnamefont {D.}~\bibnamefont {Krause}}, \bibinfo {author} {\bibfnamefont
  {S.~E.}\ \bibnamefont {Maxwell}}, \ and\ \bibinfo {author} {\bibfnamefont
  {L.~R.}\ \bibnamefont {Hunter}},\ }\href {\doibase
  10.1103/PhysRevA.61.052507} {\bibfield  {journal} {\bibinfo  {journal} {Phys.
  Rev. A}\ }\textbf {\bibinfo {volume} {61}},\ \bibinfo {pages} {052507}
  (\bibinfo {year} {2000})}\BibitemShut {NoStop}%
\bibitem [{\citenamefont {Noel}\ \emph {et~al.}(1998)\citenamefont {Noel},
  \citenamefont {Griffith},\ and\ \citenamefont {Gallagher}}]{Noel1998}%
  \BibitemOpen
  \bibfield  {author} {\bibinfo {author} {\bibfnamefont {M.~W.}\ \bibnamefont
  {Noel}}, \bibinfo {author} {\bibfnamefont {W.~M.}\ \bibnamefont {Griffith}},
  \ and\ \bibinfo {author} {\bibfnamefont {T.~F.}\ \bibnamefont {Gallagher}},\
  }\href {\doibase 10.1103/PhysRevA.58.2265} {\bibfield  {journal} {\bibinfo
  {journal} {Phys. Rev. A}\ }\textbf {\bibinfo {volume} {58}},\ \bibinfo
  {pages} {2265} (\bibinfo {year} {1998})}\BibitemShut {NoStop}%
\bibitem [{\citenamefont {Prasannaa}\ \emph {et~al.}(2015)\citenamefont
  {Prasannaa}, \citenamefont {Vutha}, \citenamefont {Abe},\ and\ \citenamefont
  {Das}}]{Prasannaa2015}%
  \BibitemOpen
  \bibfield  {author} {\bibinfo {author} {\bibfnamefont {V.~S.}\ \bibnamefont
  {Prasannaa}}, \bibinfo {author} {\bibfnamefont {A.~C.}\ \bibnamefont
  {Vutha}}, \bibinfo {author} {\bibfnamefont {M.}~\bibnamefont {Abe}}, \ and\
  \bibinfo {author} {\bibfnamefont {B.~P.}\ \bibnamefont {Das}},\ }\href
  {\doibase 10.1103/PhysRevLett.114.183001} {\bibfield  {journal} {\bibinfo
  {journal} {Phys. Rev. Lett.}\ }\textbf {\bibinfo {volume} {114}},\ \bibinfo
  {pages} {183001} (\bibinfo {year} {2015})}\BibitemShut {NoStop}%
\bibitem [{\citenamefont {Isaev}\ \emph {et~al.}(2010)\citenamefont {Isaev},
  \citenamefont {Hoekstra},\ and\ \citenamefont {Berger}}]{Isaev2010}%
  \BibitemOpen
  \bibfield  {author} {\bibinfo {author} {\bibfnamefont {T.~A.}\ \bibnamefont
  {Isaev}}, \bibinfo {author} {\bibfnamefont {S.}~\bibnamefont {Hoekstra}}, \
  and\ \bibinfo {author} {\bibfnamefont {R.}~\bibnamefont {Berger}},\ }\href
  {\doibase 10.1103/PhysRevA.82.052521} {\bibfield  {journal} {\bibinfo
  {journal} {Phys. Rev. A}\ }\textbf {\bibinfo {volume} {82}},\ \bibinfo
  {pages} {052521} (\bibinfo {year} {2010})}\BibitemShut {NoStop}%
\bibitem [{\citenamefont {{NL-eEDM collaboration: P. Aggarwal}}\ \emph
  {et~al.}(2018)\citenamefont {{NL-eEDM collaboration: P. Aggarwal}} \emph
  {et~al.}}]{Aggarwal2018}%
  \BibitemOpen
  \bibfield  {author} {\bibinfo {author} {\bibnamefont {{NL-eEDM collaboration:
  P. Aggarwal}}} \emph {et~al.},\ }\href {\doibase 10.1140/epjd/e2018-90192-9}
  {\bibfield  {journal} {\bibinfo  {journal} {Eur. Phys. J. D}\ }\textbf
  {\bibinfo {volume} {72}},\ \bibinfo {pages} {197} (\bibinfo {year}
  {2018})}\BibitemShut {NoStop}%
\bibitem [{\citenamefont {Abe}(2018)}]{Abe2018}%
  \BibitemOpen
  \bibfield  {author} {\bibinfo {author} {\bibfnamefont {M.}~\bibnamefont
  {Abe}},\ }\href@noop {} {}\bibinfo {howpublished} {private communication}
  (\bibinfo {year} {2018})\BibitemShut {NoStop}%
\end{thebibliography}%

\onecolumngrid

\newpage
\section*{Supplementary Material}

\subsection{A. Molecular orientation in an electric field}
The field-free Hamiltonian for the electronic ($^2\Sigma$) and vibrational $(v=0)$ ground state of a molecule such as $^{174}$Yb$^{107}$Ag is
\begin{equation}\label{eq:molecular_hamiltonian}
    H_0 = B_\mrm{rot} N(N+1) + \gamma \vec{S} \cdot \vec{N} + b \vec{S} \cdot \vec{I} + c S_z I_z,
\end{equation}
where $\vec{N}, \vec{S}, \vec{I}$ are the molecular rotational angular momentum, electron spin and nuclear spin respectively. $B_\mrm{rot}$ is the rotational constant of the molecule, $\gamma$ is the spin-rotation parameter, and $b,c$ are hyperfine interaction parameters. The interaction Hamiltonian with an electric field [see Eq.\ (\ref{eq:hint_1})] is $H_\mrm{int} = - D \hat{n} \cdot \Evec$, where $D$ is the molecular dipole moment. The characteristic scale of the electric field needed to polarize the molecule is $\Esca_\mrm{pol} = 2 B_\mrm{rot}/D$.

The spectroscopic constants for the YbAg molecule have yet to be measured, so for our calculations we estimated the values of $B_\mrm{rot},\gamma,b,c$ and $D$ using the measured values for the structurally very similar molecule $^{174}$Yb$^{19}$F \cite{Sauer1996}. We assume that the bond length and molecular dipole moment in YbAg are similar to that of YbF. The value of $B_\mrm{rot}$ was scaled from that of YbF by the ratio of the reduced masses of Yb-Ag and Yb-F. The values of $b$ and $c$ were scaled from the YbF values by the ratio of the nuclear magnetic moments of $^{107}$Ag and $^{19}$F. We find that $\Esca_\mrm{pol} \approx 2$ kV/cm for YbAg.

\begin{figure}[h!]
    \centering
    \includegraphics[width=0.6\textwidth]{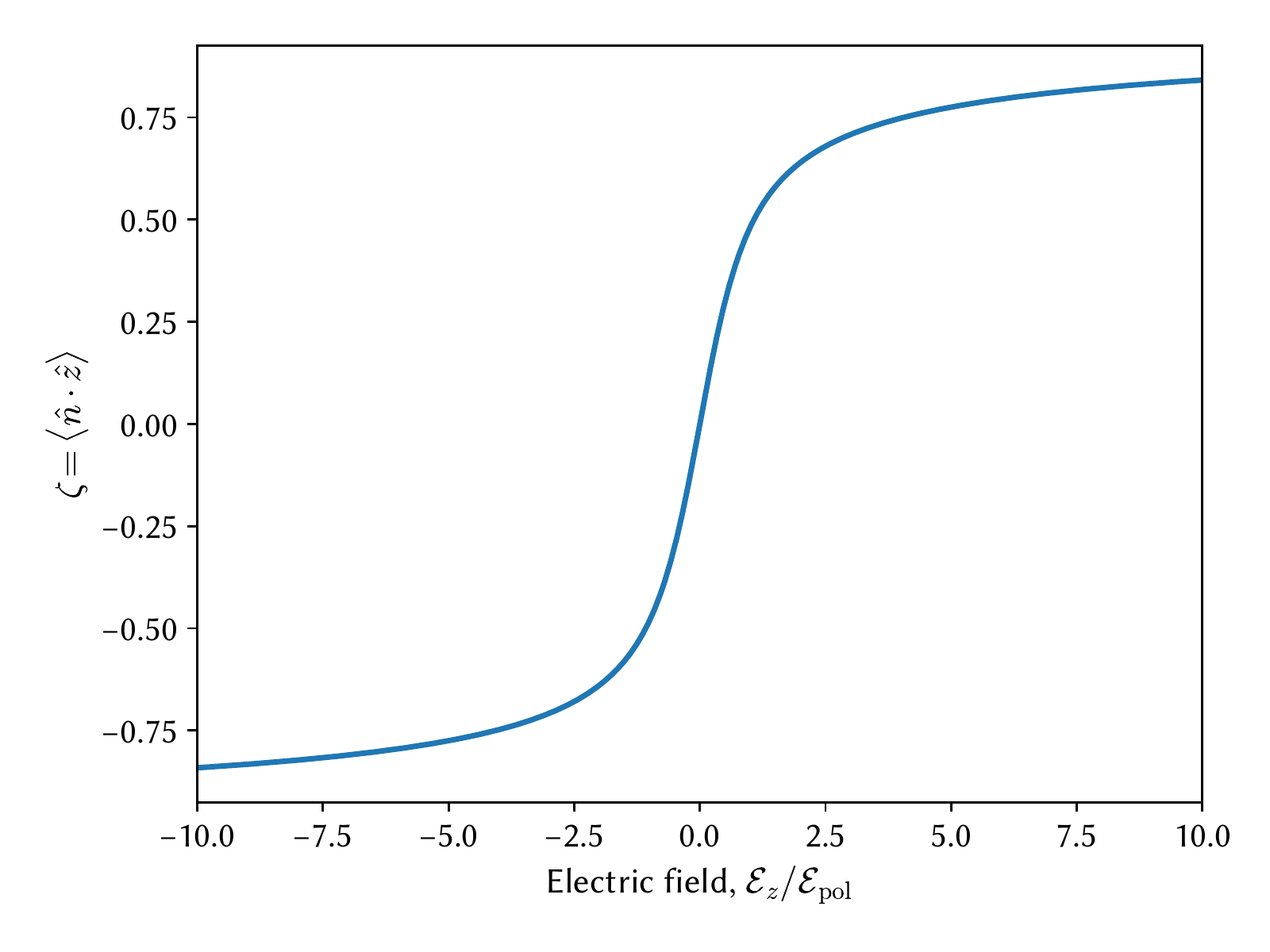}
    \caption{Molecular orientation $\zeta = \avg{\hat{n} \cdot \hat{z}}$ as a function of the electric field applied to the molecule.}
    \label{fig:zeta_vs_E}
\end{figure}

We used an uncoupled computational basis $\ket{N,m_N; S,m_S; I,m_I}$, including rotational levels up to $N=20$, and numerically diagonalized $H_0 + H_\mrm{int}$ for different values of $\Esca_z$. The resulting dependence of the molecular orientation $\zeta = \avg{\hat{n} \cdot \hat{z}}$ is shown in Fig.\ \ref{fig:zeta_vs_E}. The main features of the $\zeta$ vs. $\Esca_z$ curve can be understood from the simple approximate expression derived in Section B. When the calculated curve of $\zeta$ versus $\Esca_z$ is applied to a sinusoidal electric field $\Esca_z(t) = \Esca_0 \cos(\omega t + \beta)$ (with $\Esca_0 = 3 \Esca_\mrm{pol}$), the curve for $\zeta(t)$ shown in Fig.\ \ref{fig:Bfield_zeta_vs_time} is obtained. 

We also used the numerical model to calculate systematic errors, such as the $E1-M1$ mixing-induced Rabi frequency $\Omega_{E1-M1}$ described in the main text. For example, the numerical calculations confirm the estimate from perturbation theory, $\Omega_{E1-M1} \sim D \Esca_0 \, D \Esca_\mrm{dc} \, g_S \mu_B \Bsca_\mrm{dc} \, \frac{\gamma^2}{B_\mrm{rot}^4}$.

\begin{figure}[h!]
    \centering
    \includegraphics[width=0.6\textwidth]{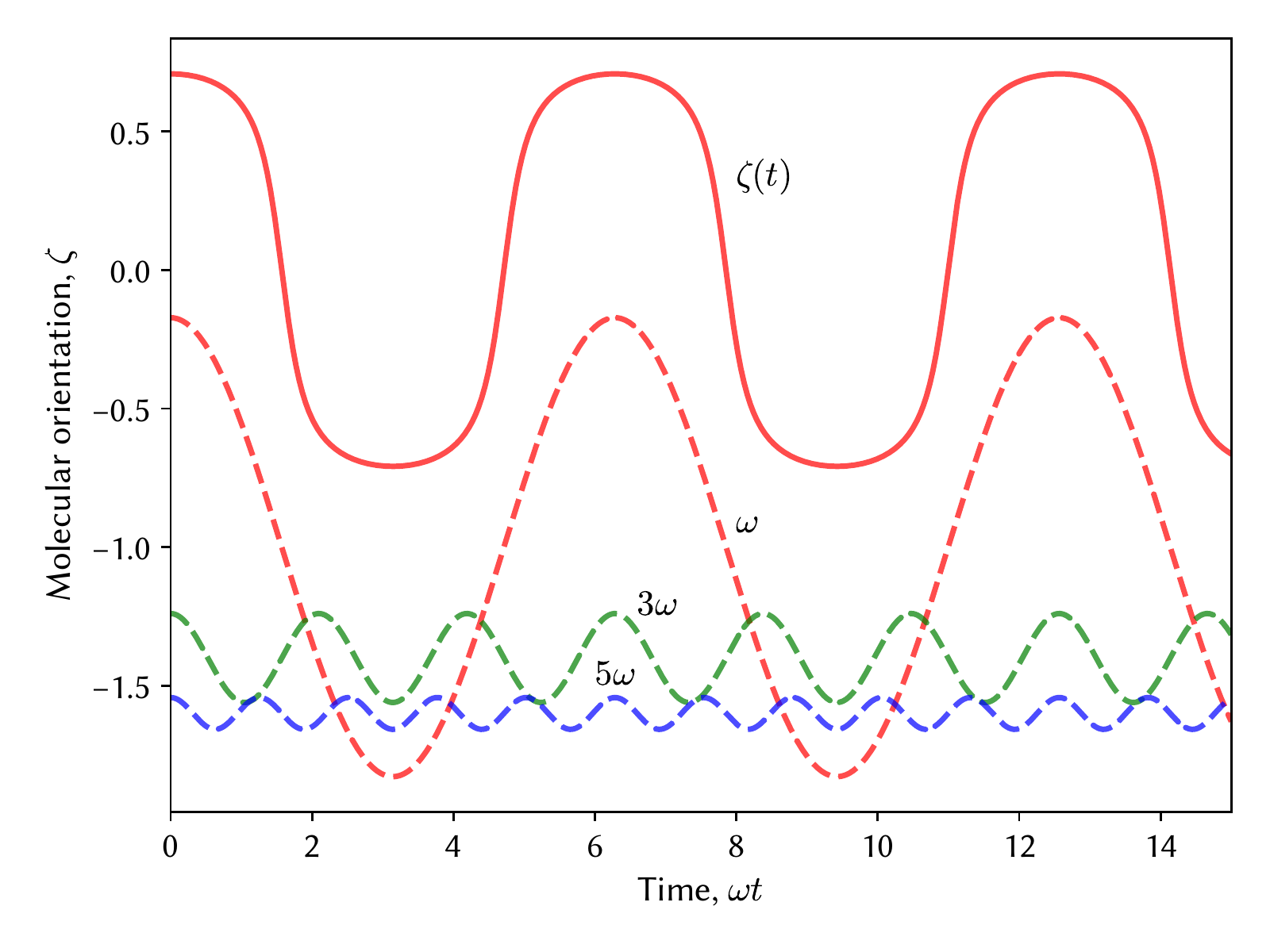}
    \caption{Molecular orientation $\zeta = \avg{\hat{n} \cdot \hat{z}}$ in response to an electric field $\Esca_z(t) = 3 \Esca_\mrm{pol} \cos \omega t$. The harmonics contained in $\zeta(t)$ are offset for clarity.}
    \label{fig:harmonics}
\end{figure}

\subsection{B. Analytical expression for $\zeta$ in a two-level system}
The quantity $\zeta$ quantifies the orientation of a polar molecule along an applied electric field. In the main text, we numerically calculate $\zeta$ from the Hamiltonian in Equation (\ref{eq:hint_1}) by considering the first 20 rotational levels. However, some useful intuition for this quantity can be gained from an approximate analytical expression for the lowest pair of (opposite parity) rotational states, $\ket{0} \equiv \ket{N=0}$ and $\ket{1} \equiv \ket{N=1,m_N=0}$. The Hamiltonian matrix for this two-level rotor system in an electric field is 
\begin{equation}
    H = \begin{pmatrix}
            -\omega_{01}/2 & -D_{01} \Esca_z \\
			 -D_{01} \Esca_z & \omega_{01}/2
			  \end{pmatrix}
\label{eq:RWA}
\end{equation}
where $D_{01} = \bra{0}D_z\ket{1}$ and $\omega_{01}$ is the spacing between $\ket{0}$ and $\ket{1}$. Typical rotational level spacing in polar molecules ($\omega_{01} \sim 2 \pi \times$ 10 GHz) are much larger than the electric field drive frequency ($\omega \sim 2\pi \times$ 100 MHz, near the hyperfine resonance frequency). Therefore, a rotating wave approximation is not valid in this regime, and a better description of the dynamics results from a quasi-static (i.e., adiabatic) approximation. Therefore, we solve for $\zeta$ using the eigenvectors of the above Hamiltonian, assuming that $\Esca_z(t)$ varies slowly compared to the phase of the wavefunction ($\omega \ll \omega_{01}$). 

The eigenvectors of the Hamiltonian are then $\ket{\widetilde{0}} = \cos(\frac{\theta}{2})\ket{0} - \sin(\frac{\theta}{2})\ket{1}$ and $\ket{\widetilde{1}} = \cos(\frac{\theta}{2})\ket{1} + \sin(\frac{\theta}{2})\ket{0}$ where $\theta = \tan^{-1}(2 D_{01}\Esca_z/\omega_{01})$. The expression for $\zeta$ in the ground state is therefore
\begin{equation}
\begin{split}
    \zeta(t) & = \frac{\bra{\widetilde{0}}D_z\ket{\widetilde{0}}}{D_{01}} = -\sin \theta = \frac{-D_{01} \Esca_z(t)}{\sqrt{[D_{01} \Esca_z(t)]^2 + (\omega_{01}/2)^2}} \\
    & \approx -\frac{2 D_{01} \Esca_0}{\omega_{01}} \cos (\omega t + \beta) \left[ 1 - \frac{1}{2} \left(\frac{2 D_{01} \Esca_0}{\omega_{01}}\right)^2 \cos^2 (\omega t + \beta) + \ldots  \right].
    \label{eq:zetalab}
\end{split}
\end{equation}
This expression also immediately shows the features of $\zeta(t)$ that are relevant to our method: (a) the amplitude of $\zeta$ approaches 1 when $D_{01} \Esca_z$ exceeds $\omega_0$, as shown in Fig.\ \ref{fig:zeta_vs_E}, and (b) the time-dependence of $\zeta(t)$ contains higher harmonics of $\omega$ because of its nonlinear dependence on $\Esca_z(t)$, as shown in Fig.\ \ref{fig:harmonics}.

\subsection{C. Suppressing systematics due to the oscillating $\Esca$-field}
We note that an electron EDM precision $\delta d_e = 10^{-31}$ $e$ cm corresponds to a measurement of the $P,T$-violating Rabi frequency $\Omega_{PT}$ with a precision of $\delta \Omega_{PT} = 2\pi \times 0.5$ $\mu$Hz. 

An oscillating $\Esca$-field in the region occupied by the molecules induces a $\Bsca$-field with amplitude $\Bsca_d \sim \frac{\ell \omega_E}{c^2} \Esca_0$, where $\ell$ is a length scale on the order of the electrode size. With $\ell \sim$ 1 cm,  $\Esca_0 \sim$ 2 kV/cm and $\omega_E = 2\pi \times$ 10 MHz, the displacement $\Bsca$-field has amplitude $\Bsca_d \sim$ 10 mG, which leads to a spurious Rabi frequency $\Omega_d \sim 2\pi \times$ 20 kHz that mimics $\Omega_{PT}$. 

This effect can be suppressed in two separate ways. First, we note that the displacement $\Bsca$-field is perpendicular to $\Evec$, which suppresses shifts in $\rho_{ee}$ because $\Bsca$-fields in the $xy$-plane only couple $\ket{F=0,m_F=0}$ to the $\ket{F'=1,m_{F'}=\pm1}$ levels. Further, the $\ket{F=0,m_F=0} \to \ket{F'=1,m_{F'}=\pm1}$ transitions are out of resonance with the frequency $\omega_E = \omega_0$ due to the tensor Stark shift in polar molecules \cite{Khriplovich1997,Hudson2002,DeMille2000}. 

Any residual $\Bsca$-field component along $\hat{z}$, for example due to electrode mis-alignment, can be suppressed using a second mechanism. Note that the induced $\Bsca$-field is proportional to the time derivative of $\Esca_z$, and so it lags the applied $\Esca$-field in phase by $\pi/2$. For example, if $\Bvec_d \cdot \hat{z} \neq 0$ and $\beta$ is set to $\pm \frac{\pi}{2}$, the change in $\rho_{ee}$ depends on whether the $\Esca$-field is on or off. Such a shift is \textit{only} produced when $\Bvec_d \cdot \hat{z} \neq 0$, and is therefore a clean diagnostic for displacement $\Bsca$-fields. 

Despite these two suppression mechanisms, it is possible that a combination of phase errors (e.g., due to charging currents or cable impedance mismatches) and electrode misalignments could lead to a residual $\Bsca$-field that is both parallel to $\hat{z}$ and in phase with $\Esca_z$. For example misalignment of the rf electric and magnetic field directions by $\theta$ = 10$^{-5}$ rad, and a phase error in the electric field drive of $\Delta \beta$ = 10$^{-5}$ rad will lead to a residual Rabi frequency $\Omega_d' \sim \theta \ \Delta \beta \ \Omega_d = 2\pi \times 10^{-6}$ Hz. This small error can be detected and further suppressed using the sub-harmonic modulation method described next.

%\subsection{D. Sub-harmonic modulation: a diagnostic for systematic errors}
%
As noted in the main text, the molecular orientation $\zeta$ is \emph{nonlinear} in $\Esca_z$. If $\Esca_0$ is large enough to appreciably polarize the molecule, then $\zeta(t)$ also contains higher odd harmonics of $\omega_E$ (see Fig.\ \ref{fig:harmonics}). This fact leads to a unique and powerful diagnostic for systematic errors: an EDM search experiment can be conducted using, e.g., $\omega_E = \omega_0/3$ and $\omega_B = \omega_0$. Any induced magnetic fields that are linear in $\Esca_z$ (a condition which covers the majority of conceivable systematics) oscillate at $\omega_0/3$, far off resonance from the clock transition, and so their interference with the transition amplitude is greatly suppressed. On the other hand, the fourier component of $\zeta$ that oscillates at $3\omega_E = \omega_0$ (with amplitude $\zeta_3$) resonantly contributes to the transition probability as $\rho_{ee}(\tau) = \sin^2\left[\frac{(\Omega_B + \Omega_{PT,3} \, \cos 3 \beta)\,\tau}{2} \right]$, with $\Omega_{PT,3} = \frac{1}{2} W_{PT} \zeta_3$. Therefore, driving the electric field at a sub-harmonic of $\omega_0$ offers a convenient diagnostic to discriminate between systematic errors and real $P,T$-violating signals. Since $\zeta_3 < \zeta_0$ though, it yields lower EDM sensitivity, so we envision that experiments using the proposed method will intersperse some measurements with $\omega_E = \omega_0/3$ as systematic checks within larger measurement blocks with $\omega_E = \omega_0$. 

\subsection{D. Differential Stark shift}

Under the influence of the electric field $\Esca_z(t) = \Esca_0 \cos(\omega t + \beta)$, there is a small differential Stark shift (DSS) between the hyperfine clock states $\ket{g}, \ket{e}$. This shift arises due to a combination of the $E1$ interaction of the molecular dipole moment with the electric field, the spin-rotation interaction, and the hyperfine interaction. From perturbation theory, we find that the resulting change in the hyperfine resonance frequency $\omega_0$ is 
\begin{equation}
\Delta \omega_\mrm{DSS} \sim \frac{\gamma^2 \, b}{B_\mrm{rot}^4} \, (D \Esca_0)^2  \cos^2(\omega t + \beta).
\end{equation}
We have also confirmed this expression with direct numerical calculations using the Hamiltonian in Equation \ref{eq:molecular_hamiltonian}. 

The static part of $\Delta \omega_\mrm{DSS}$ can be absorbed into the definition of the resonance frequency $\omega_0$, leaving a modulation of the hyperfine splitting at a frequency $2 \omega$. Therefore the effect of the DSS can be described by writing the resonance frequency as $\omega_0' = \omega_0 \, [1 + h \cos(2\omega t + 2 \beta)]$, where $h$ is a dimensionless parameter. The numerically calculated value of $h$ is $\sim 10^{-6}$ at $\Esca_0 \sim$ 2 kV/cm for YbAg. It now remains to consider the effect of such a modulation of the resonance frequency on the time-evolution of $\rho_{ee}(t)$. 

The Rabi problem for a two-level system with a modulated resonance frequency has not appeared in the literature to the best of our knowledge (although the Rabi problem with a frequency-modulated drive frequency is well-understood, cf. \cite{Noel1998}). Therefore we briefly describe the method of solution here. 

We work in the interaction picture, wherein the magnetic dipole moment operator is represented by the matrix 
\begin{equation}
\mu_z = \begin{pmatrix}
0 & \mu_{ge} \, e^{-i \int \omega_0' \, dt} \\ 
\mu_{eg} \, e^{+i \int \omega_0' \, dt} & 0\\ 
\end{pmatrix} = \begin{pmatrix}
0 & \mu_{ge} \, e^{-i \omega_0 t} \, e^{+ i \frac{h \omega_0}{2 \omega} \sin(2 \omega t + 2 \beta)}\\ 
\mu_{eg} \, e^{+i \omega_0 t} \, e^{- i \frac{h \omega_0}{2 \omega} \sin(2 \omega t + 2 \beta)} & 0\\ 
\end{pmatrix}
\end{equation}
in the two-level subspace spanned by $\ket{g},\ket{e}$. The Hamiltonian for the hyperfine transition is 
\begin{equation}
H_\mrm{hf}' = -\mu_z \Bsca_0 \cos \omega t  = \Omega_B \begin{pmatrix}
0 & e^{-i \omega_0 t} \, e^{+ i \frac{h \omega_0}{2 \omega} \sin(2 \omega t + 2 \beta)} \\ e^{+i \omega_0 t} \, e^{- i \frac{h \omega_0}{2 \omega} \sin(2 \omega t + 2 \beta)} & 0 \end{pmatrix} \cos \omega t.
\end{equation}

Following the usual approach to the rotating-wave approximation, we retain just the slowest terms in this Hamiltonian. We also expand the phase modulation term up to $\mathcal{O}(h^2)$ owing to the smallness of $h$, and get
\begin{equation}
H_\mrm{hf}' \approx \frac{\Omega_B}{2} \begin{pmatrix}
0 & e^{+i \Delta t} \, \left[1 + \left(\frac{h\omega_0}{4 \omega}\right) \, e^{i 2 \beta}\right] \\ e^{-i \Delta t} \, \left[1 + \left(\frac{h\omega_0}{4 \omega}\right) \, e^{-i 2 \beta}\right] & 0 \end{pmatrix} = \frac{1}{2} \begin{pmatrix}
0 &  \Omega_M \, e^{+i\Delta t} \\ \Omega_M^* \, e^{-i \Delta t} & 0 \end{pmatrix}.
\end{equation}

This has exactly the same form as the Hamiltonian for the standard Rabi problem with a fixed hyperfine splitting, $H_\mrm{hf} = \frac{1}{2} \begin{pmatrix}
0 &  \Omega_B \, e^{+i\Delta t} \\ \Omega_B^* \, e^{-i \Delta t} & 0 \end{pmatrix}$. Importantly, this means $\rho_{ee}(t)$ evolves in time in just the same way as in the unmodulated case. The only difference is that the Rabi frequency is modified to $\Omega_M = \Omega_B \left[1 + \left(\frac{h\omega_0}{4 \omega}\right) \, e^{i 2 \beta}\right]$. Direct numerical solutions of the Schrodinger equation with a modulated resonance frequency confirm this simple picture. 

The shift in the measured Rabi frequency on resonance is then $\Omega_\mrm{DSS} = \frac{h}{4}  \Omega_B\, \cos2 \beta$. With $\Omega_B = \pi/2 \tau = 2 \pi \times 25$ mHz and $h=10^{-6}$, this evaluates to $\Omega_\mrm{DSS} \sim$ $2 \pi \times 6$ nHz $\cos 2\beta$, which is an extremely small effect compared to the targeted precision $\delta \Omega_{PT} \sim 2 \pi \times$ 500 nHz.

Nevertheless, we show how it can be suppressed further. Note that due to its $\cos 2 \beta$ dependence, this shift is the same for $\beta=0$ and $\beta=\pi$, whereas the $PT$-violating observable $\Omega_{PT} \cos \beta$ switches sign between these two phases. Further note that $\Omega_\mrm{DSS} \propto \Omega_B \propto \Bsca_0$. Therefore measurements of $\rho_{ee}(\tau)$ with different values of $\Omega_B \tau$ (say $\frac{\pi}{2}$ and $\frac{5 \pi}{2}$) can distinguish a genuine $PT$-violating signal from DSS-induced transition amplitudes. These, in combination with the other diagnostics described above (e.g., sub-harmonic drive, electric field amplitude variation) at our disposal, lead us to conclude that DSS effects can be cleanly distinguished from a true $P,T$-violating signal.

\subsection{E. A menu of molecules for EDM searches}
As an illustration of the variety of polar molecules to which our measurement method can be applied, the following tables list neutral molecules (Table I) and singly-charged molecular ions (Table II) that can be used to search for electron and nuclear EDMs with our method. In each table, a combination of the atoms  from the two columns forms an EDM-sensitive molecule to which our method can be applied. Atoms that can be cooled to ultracold temperatures are shown in bold: ultracold assembled molecules can be produced from pairs of these.

The tables include molecules that have been previously used (YbF \cite{Hudson2011}) or proposed for use in EDM experiments (HgF, HgCl, HgBr \cite{Prasannaa2015}, HgNa, HgK, HgRb \cite{Sunaga2018}, RaF \cite{Isaev2010}, BaF \cite{Aggarwal2018}, RaAg \cite{Fleig_DeMille}, HgCa \cite{Abe2018}). The tables are by no means exhaustive -- our technique can be used with other molecules too (e.g., TlF, $^{225}$Ra$^{199}$Hg). 

\begin{table}[h!]
    \large
    \caption{Neutral molecules.}
    \begin{minipage}{.3\linewidth}
      \caption*{Electron EDM}
        \begin{tabular}{|c|c|}
            \hline
            $^{138}$\textbf{Ba} & $^{19}$F \\ 
            $^{174}$\textbf{Yb} & $^{35}$Cl \\ 
            $^{202}$\textbf{Hg} &  $^{79}$Br \\
            $^{226}$\textbf{Ra} & $^{107, 109}$\textbf{Ag}  \\
             ~         & $^{16}$O$^{1}$H \\ 
            \hline
        \end{tabular}
    \end{minipage}%
    \begin{minipage}{.3\linewidth}
        \caption*{Nuclear EDM}
        \begin{tabular}{|c|c|}
            \hline
            $^{199}$\textbf{Hg} & $^{17}$O \\ 
            $^{207}$Pb & $^{33}$S \\ 
            $^{225}$\textbf{Ra} & $^{43}$\textbf{Ca} \\
                     ~ & $^{87}$\textbf{Sr} \\
            \hline
        \end{tabular}
    \end{minipage} 
\end{table}

\begin{table}[h!]
\large
    \caption{Molecular ions.}
    \begin{minipage}{.3\linewidth}
      \caption*{Electron EDM}
      \begin{tabular}{|c|c|}
            \hline
            $^{200}$Hg & $^{17}$O \\ 
            $^{226}$Ra & $^{33}$S \\ 
            $^{208}$Pb & $^{43}$Ca \\
            $^{232}$Th & $^{87}$Sr \\
            \hline
        \end{tabular}
    \end{minipage}%
    \begin{minipage}{.3\linewidth}
        \caption*{Nuclear EDM}
        \begin{tabular}{|c|c|}
            \hline
            $^{133}$Ba & $^{19}$F \\ 
            $^{199}$Hg & $^{35}$Cl \\ 
            $^{207}$Pb & $^{79}$Br \\
            $^{225}$Ra & $^{16}$O$^{1}$H \\
            $^{229}$Pa & \\
            $^{229}$Th & \\
            \hline
        \end{tabular}
    \end{minipage} 
\end{table}

 ~ ~ ~ ~\\ ~ ~ ~

\end{document}